\documentclass[a4paper,fleqn,usenatbib]{mnras}

\usepackage{newtxtext,newtxmath}
\usepackage[T1]{fontenc}
\usepackage{ae,aecompl}

\usepackage{graphicx}	% Including figure files
\usepackage{amsmath}	% Advanced maths commands
\usepackage{amssymb}	% Extra maths symbols
\usepackage{xcolor, comment, epstopdf, ulem, mathtools, amsfonts}

\newcommand{\rmd}{\,\mathrm{d}}
\newcommand{\rmi}{\,\mathrm{i}}
\newcommand{\del}{\partial}

\title[Mixed modes with moderate core fields]{Low-degree mixed modes in red giant stars with moderate core magnetic fields}

\author[Loi \& Papaloizou]{
Shyeh Tjing Loi\thanks{E-mail: stl36@cam.ac.uk} and
John C. B. Papaloizou\thanks{E-mail: jcbp2@damtp.cam.ac.uk}
\\
Department of Applied Mathematics and Theoretical Physics, University of Cambridge, Centre for Mathematical Sciences, Wilberforce Road, Cambridge CB3 0WA, UK
}

\date{Accepted XXX. Received YYY; in original form ZZZ}

\pubyear{2019}

\begin{document}
\label{firstpage}
\pagerange{\pageref{firstpage}--\pageref{lastpage}}
\maketitle

% Abstract of the paper
\begin{abstract}
  Observations of pressure-gravity mixed modes, combined with a theoretical framework for understanding mode formation, can yield a wealth of information about deep stellar interiors. In this paper, we seek to develop a formalism for treating the effects of deeply buried core magnetic fields on mixed modes in evolved stars, where the fields are moderate, i.e.~not strong enough to disrupt wave propagation, but where they may be too strong for non-degenerate first-order perturbation theory to be applied. The magnetic field is incorporated in a way that avoids having to use this. Inclusion of the Lorentz force term is shown to yield a system of differential equations that allows for the magnetically-affected eigenfunctions to be computed from scratch, rather than following the approach of first-order perturbation theory. For sufficiently weak fields, coupling between different spherical harmonics can be neglected, allowing for reduction to a second-order system of ordinary differential equations akin to the usual oscillation equations that can be solved analogously. We derive expressions for (i) the mixed-mode quantisation condition in the presence of a field and (ii) the frequency shift associated with the magnetic field. In addition, for modes of low degree we uncover an extra offset term in the quantisation condition that is sensitive to properties of the evanescent zone. These expressions may be inverted to extract information about the stellar structure and magnetic field from observational data.
\end{abstract}

% Select between one and six entries from the list of approved keywords.
% Don't make up new ones.
\begin{keywords}
  MHD -- methods: analytical -- stars: interiors -- stars: magnetic field -- stars: oscillations
\end{keywords}

%%%%%%%%%%%%%%%%%%%%%%%%%%%%%%%%%%%%%%%%%%%%%%%%%%

%%%%%%%%%%%%%%%%% BODY OF PAPER %%%%%%%%%%%%%%%%%%

\section{Introduction}
The ability to infer stellar properties based on the observed frequencies of oscillation relies on a theoretical understanding of the conditions under which modes form. An oscillation mode arises from the constructive interference of propagating waves, of which there are various types: acoustic waves give rise to p-modes, while gravity waves give rise to g-modes, these being restored by pressure and buoyancy, respectively. An asymptotic analysis predicts p-modes to be spaced evenly in frequency, and g-modes to be evenly spaced in period \citep[e.g.][]{Gough1993}. The frequency/period spacings and patterns of deviation from constant spacing depend on the depth profiles of the sound speed $c_s$ and the buoyancy frequency $N$. Analysis of the seismic spectrum thus yields information about the internal structure and properties of a star \citep{Stello2009, Huber2010, Chaplin2011, Aguirre2013, Yu2018}.

\begin{figure*}
  \centering
  \includegraphics[width=0.9\textwidth]{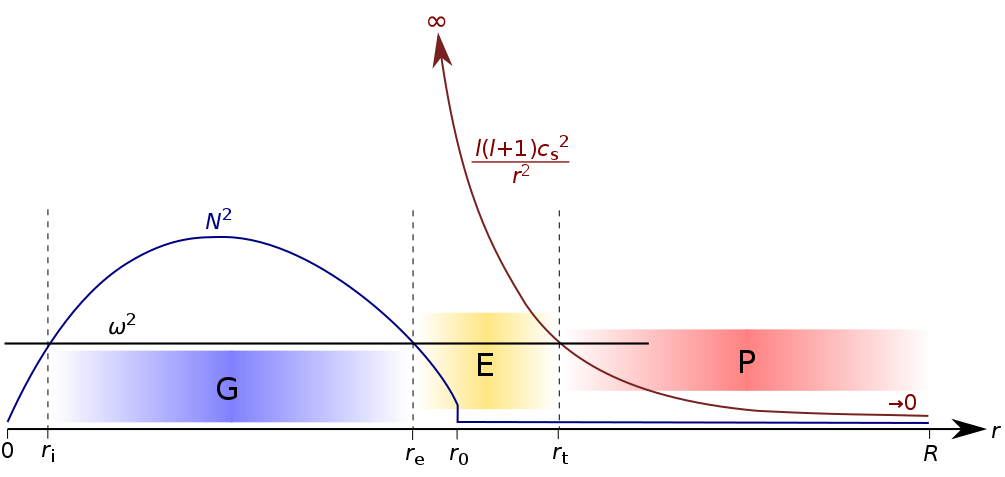}
  \caption{Schematic diagram (note: not to scale) showing the labelling of the different regions of the model. Region G corresponds to the g-mode cavity, where $\omega^2 < N^2$, and is bounded by turning points at radial coordinates of $r_i$ and $r_e$, corresponding to where $\omega^2 = N^2$. Region P corresponds to the p-mode cavity, where $\omega^2 > \ell(\ell+1) c_s^2 / r^2$, and extends from near the surface to an inner turning point at radial coordinate $r_t$, where $\omega^2 = \ell(\ell+1) c_s^2 / r^2$. Region E is the evanescent zone, lying between $r_e$ and $r_t$. $N^2$ is taken to be zero in all of P and part of E, dropping to zero at a radial coordinate of $r_0$. The sound speed $c_s^2$ is assumed to be infinite in G. See Section \ref{sec:hydro_bgs} for further discussion of these regions.}
  \label{fig:RG_regions}
\end{figure*}

In main sequence stars, the cavities to which p-modes and g-modes of a given frequency localise are well separated in radius, and their mutual coupling is negligible. However, when a star evolves off the main sequence, the radial distance between the two cavities shrinks, allowing the two types of fluid motion to couple. This gives rise to \textit{mixed modes} having both p- and g-like character. Mixed modes are useful for probing the properties of the deep radiative interior (where g-modes are supported), since the acquisition of p-like character enhances their surface amplitudes and therefore observability. Observations of mixed modes have led to numerous breakthroughs in characterising the internal structure of red giants, including the ability to measure core rotation rates \citep{Beck2011, Deheuvels2012} and distinguish between H shell- and He core-burning stages \citep{Bedding2011}. The possibility of mixed modes was predicted much earlier, first in numerical studies by \citet{Osaki1975} and \citet{Aizenman1977}, which were then followed up in analytical work by \citet{Shibahashi1979} and \citet{Tassoul1980}. These analytical studies made use of the Cowling approximation and the assumption of a thick evanescent zone separating the two cavities, but the more recent work of \citet{Takata2016, Takata2016a} extended this to the case of a thin evanescent zone (strong coupling) and removing the Cowling approximation for the special case of dipole modes. Most work on mixed modes has been numerical rather than analytical; however, analytical results for mixed modes have pertinent application to studying various aspects of red giant structure, for example, measuring properties in the vicinity of the evanescent zone \citep{Mosser2012a, Mosser2017, Hekker2018, Pincon2019}.

Alongside the successes in probing rotation and buoyancy properties of the core, a more recent result surrounds the attribution of depressed amplitudes for low-degree mixed modes in about 20\% of the red giant population to strong (dynamically significant) core magnetic fields \citep{Fuller2015}. Follow-up work investigating the observational consequences applied non-degenerate perturbation theory to estimating frequency splittings, and found these to be comparable to g-mode spacings \citep{Cantiello2016}. More recent work has demonstrated that the process of mode formation under strong fields is likely to be subject to dynamical chaos \citep{Loi2018}, and hence far removed from the paradigms of the standard theory of stellar oscillations, under which effects such as magnetism can be treated perturbatively \citep{Gough1990}. However, even considering fields that are small enough to avoid the chaotic regime, the results of non-degenerate perturbation theory should be questioned when frequency shifts become comparable to mode spacings. In this scenario, degenerate perturbation theory would more appropriately be employed \citep[e.g.][]{Soufi1998, Kiefer2017, Kiefer2018}. This involves treating the unperturbed eigenfunctions as superpositions of near-degenerate modes, whose coefficients must be obtained by a matrix inversion. This is one possible approach to dealing with the effects of magnetism in the moderate regime. By \textit{moderate}, we mean where field strengths are not large enough to be dynamically significant (potentially leading to mode damping/chaos), but are large enough to produce frequency shifts of the order/in excess of underlying mode spacings and thus cannot be adequately described by non-degenerate perturbation theory. For short-wavelength modes, perturbation theory predicts a shift of the order $\Delta \omega \sim k^2 v_A^2/\omega$, where $k$ is the angular wavenumber, $v_A = B/\sqrt{\mu_0 \rho}$ is the Alfv\'{e}n speed, $B$ is the field strength, $\mu_0$ is the vacuum permeability, $\rho$ is the mass density and $\omega$ is the angular frequency. The g-mode spacing is of the order $\Delta \omega \sim \omega^2/[\sqrt{\ell(\ell+1)} N]$, where $\ell$ is the spherical harmonic degree. Inserting typical red giant parameters of $\omega/2\pi \sim 100$\,$\upmu$Hz, $N/2\pi \sim 10$\,mHz, $k/2\pi \sim 10^{-6}$\,m$^{-1}$ and $\rho \sim 10^7$\,kg\,m$^{-3}$ suggests that the transition from weak to moderate fields occurs at about $B \sim 10^5$\,G in the case of low-degree modes. On the other hand, the transition from moderate to strong fields occurs when $\omega \sim k v_A$, corresponding to $B \sim 10^6$\,G.

The goal here is to develop a formalism for treating the oscillations of strongly coupled p- and g-cavities, where a magnetic field of moderate strength resides within the g-cavity (core region). This paper is organised as follows. In Section \ref{sec:model}, we introduce the red giant model and magnetic field configuration. In Section \ref{sec:mixedmodes}, we derive the eigenfrequency condition for mixed modes in the case of no field, then in Section \ref{sec:Bfield} we derive the modification to this resulting from the presence of a core field. Broader implications, including assumptions and limitations, are discussed in Section \ref{sec:discussion}. Finally, we conclude in Section \ref{sec:summary}.

\section{Basic equations governing oscillations of a spherical star}
Considering a star to be a self-gravitating body of fluid, its behaviour can be described using the equations of fluid dynamics, which govern how gas moves under action of various internal and external influences. Besides the forces arising from pressure and gravity, additional forces (arising from effects such as rotation and/or magnetism) may be present; collecting them into a single body force term $\mathbf{f}$, the equations governing the dynamics of a fluid are \citep[e.g.][]{Unno1989}
\begin{align}
  \frac{\del \rho}{\del t} &= -\nabla \cdot (\rho \mathbf{u}) \:, \label{eq:conty} \\
  \rho \left( \frac{\del \mathbf{u}}{\del t} + \mathbf{u} \cdot \nabla \mathbf{u} \right) &= -\nabla p - \rho \nabla \Phi + \mathbf{f} \:, \label{eq:momentum}
\end{align}
which arise from the conservation of mass and momentum, respectively. Here $\mathbf{u}$ is the fluid velocity, $p$ is the pressure, and $t$ is time. The fluid also needs to obey $\nabla^2 \Phi = 4\pi G\rho$, which is Poisson's equation for the gravitational potential $\Phi$. An equation of state, usually taken to be the adiabatic condition
\begin{align}
  \frac{\mathrm{D}p}{\mathrm{D}t} = \frac{\Gamma_1 p}{\rho} \frac{\mathrm{D}\rho}{\mathrm{D}t} \:, \\
  \shortintertext{where $\Gamma_1 = (\del \ln p/\del \ln \rho)_\text{ad}$ is the first adiabatic exponent and}
  \frac{\mathrm{D}}{\mathrm{D}t} \equiv \frac{\del}{\del t} + \mathbf{u} \cdot \nabla \:,
\end{align}
which is valid in the limit of short-timescale motions (compared to the Kelvin-Helmholtz timescale), closes the system. For small-amplitude, time-harmonic disturbances of frequency $\omega$ about a static, spherically symmetric background, we expand each scalar quantity $q \in \{ p, \rho, \Phi\}$  and fluid displacement vector $\boldsymbol{\xi}$ (satisfying $\mathbf{u} = \del \boldsymbol{\xi}/\del t$) according to
\begin{align}
  q(\mathbf{r}, t) &= q_\text{bg} (r) + q'(r) Y_\ell^m(\theta, \phi) \exp(-\rmi \omega t) \:, \\
  \boldsymbol{\xi}(\mathbf{r}, t) &= \left[ \xi_r(r) Y_\ell^m(\theta, \phi) \hat{\mathbf{r}} + \boldsymbol{\xi}_\text{h}(\mathbf{r}) \right] \exp(-\rmi \omega t) \:,
\end{align}
where $(r, \theta, \phi)$ are spherical polar coordinates, subscript `bg' refers to the background, $\hat{\mathbf{r}}$ is the radial unit vector, $\xi_r$ and $\boldsymbol{\xi}_\text{h}$ are the radial and horizontal (orthogonal to $\hat{\mathbf{r}}$) components of $\boldsymbol{\xi}$, primes indicate Eulerian perturbations, and $Y_\ell^m$ is the spherical harmonic of degree $\ell$ and order $m$. Substituting and linearising, and neglecting $\Phi'$ under the Cowling approximation \citep{Cowling1934}, which is valid when the spatial scale of oscillations is small, we arrive at the equations governing stellar oscillations,
\begin{align}
  \frac{\rmd p'}{\rmd r} &= \rho_\text{bg} \left( \omega^2 - N^2 \right) \xi_r + p' \frac{1}{p_\text{bg} \Gamma_1} \frac{\rmd p_\text{bg}}{\rmd r} \:, \label{eq:osc-1} \\
  \frac{1}{r^2} \frac{\rmd}{\rmd r} \left( r^2 \xi_r \right) &= \frac{p'}{\rho_\text{bg} c_s^2} \left( \frac{\ell(\ell+1) c_s^2}{r^2 \omega^2} - 1 \right) - \xi_r \frac{1}{p_\text{bg} \Gamma_1} \frac{\rmd p_\text{bg}}{\rmd r} \:, \label{eq:osc-2} \\
  \shortintertext{where}
  N &\equiv \left[ g \left( \frac{1}{\Gamma_1 p_\text{bg}} \frac{\rmd p_\text{bg}}{\rmd r} - \frac{1}{\rho_\text{bg}} \frac{\rmd \rho_\text{bg}}{\rmd r} \right) \right]^{1/2}
\end{align}
is the buoyancy frequency, $g$ is the gravitational acceleration, $c_s = (\Gamma_1 p_\text{bg}/\rho_\text{bg})^{1/2}$ is the adiabatic sound speed, and for the moment we set $\mathbf{f} = \mathbf{0}$. When boundary conditions are imposed, the solutions to Equations (\ref{eq:osc-1})--(\ref{eq:osc-2}) correspond to allowed eigenmodes of the star. 

\section{Red giant model}\label{sec:model}
\subsection{Hydrodynamic backgrounds}\label{sec:hydro_bgs}
In our model we will assume that $\Gamma_1$ is a constant and equal to $\gamma$. We subdivide the star into three regions, labelled G, P and E, as illustrated in Figure \ref{fig:RG_regions}. Constructing the mixed-mode quantisation condition involves deriving the solutions (which are functions of frequency) valid in each of the different regions, and identifying the condition under which they connect smoothly into one another. The description of each region is as follows:
\begin{itemize}
  \item G --- the g-mode cavity, where $\omega^2 < N^2$, corresponding to the propagation region for gravity waves. Here we assume that $c_s^2 \to \infty$, and that the spatial scale of oscillations is small which validates a Wentzel-Kramers-Brillouin-Jeffreys (WKBJ) approach, except near $r_i$ and $r_e$, which are the interior and exterior turning points, respectively.
  \item P --- the p-mode cavity, where $\omega^2 > \ell(\ell+1) c_s^2 / r^2$, corresponding to the propagation region for acoustic waves. Here we assume that $N^2 = 0$, and likewise that the spatial scale of oscillations is small validating a WKBJ approach, except near the lower turning point $r_t$ and upper turning point near the surface.
  \item E --- the evanescent zone, lying between $r_e$ and $r_t$. For small $\ell$ it is not expected that WKBJ will hold.
\end{itemize}

Without a magnetic field, the treatment of the different regions and process of matching them together is a reasonably standard procedure. This has previously been explored in the literature \citep{Shibahashi1979, Tassoul1980, Takata2016a}. In this paper we consider the possibility of a magnetic field that is confined below $r = r_e$, i.e.~residing in the g-mode cavity (region G). The way in which we incorporate this is non-standard, involving modification to the g-mode phase integral. Although three regions have been introduced in this section, the treatment of regions P and E (which are less relevant to the magnetic problem, but included for completeness) will be dealt with in Appendix \ref{sec:convectiveenvelope}, and only region G will be discussed in the main body of this paper (see Section \ref{sec:mixedmodes}).

\subsection{Magnetic field}\label{sec:Prendergast}
A red giant star may harbour a magnetic field in its core if such a field was generated by a dynamo when the star was on the main sequence, and this was able to adopt a stable configuration following cessation of the dynamo. Core dynamos are thought to operate in main sequence stars more massive than about 1.2\,M$_\odot$, corresponding to spectral types F and earlier, whose cores are convective as opposed to radiative \citep{Brun2005, Featherstone2009}. Numerical studies investigating the evolution of initially random fields have found that these tend to settle into large-scale, axisymmetric, twisted-torus configurations, with poloidal and toroidal components of comparable strength \citep{Braithwaite2004, Braithwaite2006}. The long timescales of buoyant rise of global-scale field structures imply that such fields originating in the deep interior are unlikely ever to be observed at the surface.

\begin{figure*}
  \centering
  \includegraphics[width=0.9\textwidth]{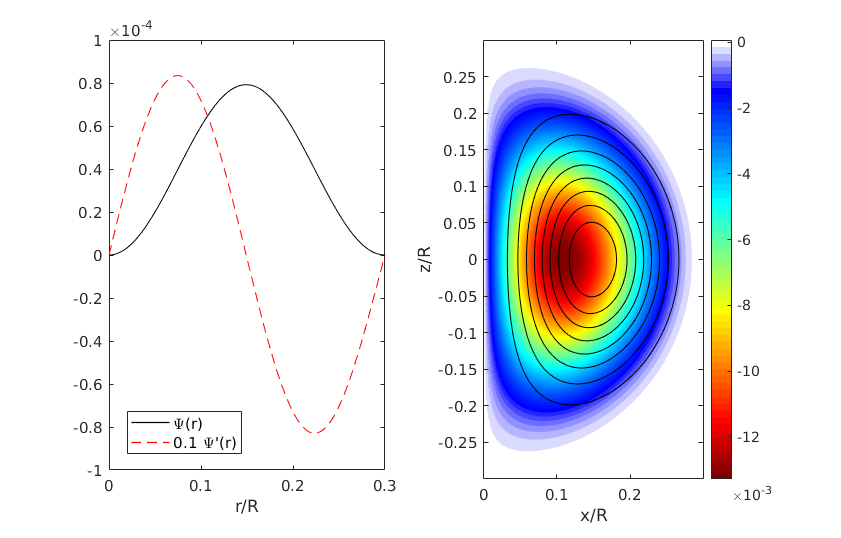}
  \caption{The Prendergast solution, computed for the sake of illustration using a solar-mass polytropic background model of index 3, where the magnetic field is confined to $r < r_f = 0.3$ in units of the stellar radius. Left: the radial flux function $\Psi$ and its derivative (scaled down by a factor of 10 for convenience of visualisation). Right: the toroidal component of the field, $B_\phi$, in colour, overlaid with a selection of poloidal field lines (level surfaces of $\Psi$). For this model, $\lambda = 21.969$, and the overall scaling is such that the central Alfv\'{e}n speed is 0.01\,$\sqrt{GM/R}$.}
  \label{fig:Prendergast_model}
\end{figure*}

An analytic expression for a spherically confined, twisted-torus field configuration was discovered by \citet{Prendergast1956} for incompressible stars, and this was extended to the compressible case by \citet{Duez2010a}. The result, which we refer to as the \textit{Prendergast solution}, is given by
\begin{align}
  &\psi(r,\theta) \propto \frac{\lambda r \sin^2 \theta}{j_1(\lambda r_f)} \left[ f_\lambda(r, r_f) \int_0^r \rho \xi^3 j_1(\lambda \xi) \rmd \xi \right. \nonumber \\
  & \qquad \left. + j_1(\lambda r) \int_r^{r_f} \rho \xi^3 f_\lambda(\xi, r_f) \rmd \xi \right] \:, \label{eq:Prendergast} \\
  \shortintertext{where $\psi$ is the poloidal flux function and $\lambda$ satisfies}
  &\int_0^{r_f} \rho \xi^3 j_1(\lambda \xi) \rmd \xi = 0 \:,
\end{align}
$r = r_f$ is the boundary of the field region, $f_\lambda(r_1, r_2) \equiv j_1(\lambda r_2) y_1(\lambda r_1) - j_1(\lambda r_1) y_1(\lambda r_2)$, and $j_1, y_1$ are spherical Bessel functions. Exterior to $r_f$ we set $\mathbf{B} = \mathbf{0}$. For an overview of how the Prendergast solution is derived, see \citet{Loi2017}. Note that in general, an axisymmetric field configuration can be expressed in terms of a poloidal flux function $\psi = \psi(r,\theta)$ by
\begin{align}
  \mathbf{B}(r,\theta) = \frac{1}{r \sin \theta} \nabla \psi \times \hat{\boldsymbol{\phi}} + B_\phi \hat{\boldsymbol{\phi}} \:.
\end{align}
In the special case of the Prendergast solution, $B_\phi$ is a function of $\lambda$ and $\Psi$, and radial and latitudinal dependencies separate according to $\psi(r,\theta) = \Psi(r) \sin^2 \theta$. The components of the field can then be expressed
\begin{align}
  \mathbf{B}(r,\theta) = \left( \frac{2}{r^2} \Psi(r) \cos \theta \:,\; -\frac{1}{r} \Psi'(r) \sin \theta \:,\; -\frac{\lambda}{r} \Psi(r) \sin \theta \right) \:,\; \label{eq:Bcompts}
\end{align}
where a prime denotes a derivative with respect to $r$. The Prendergast solution for a simple stellar model is illustrated in Figure \ref{fig:Prendergast_model}. It is an axisymmetric, dipole-like configuration, but unlike a dipole has no singularity, possesses a toroidal component in addition to the poloidal one, and vanishes smoothly at the radius $r_f$ in all three components of the field, which avoids unphysical current sheets at the boundary.

\section{Formation of mixed modes}\label{sec:mixedmodes}
The purpose of this section is to derive a condition determining the eigenfrequencies in the absence of a magnetic field. The final result is given in Equation (\ref{C8toC93}). Modifications to this arising from a core magnetic field will be dealt with subsequently in Section \ref{sec:Bfield}. Obtaining the frequency quantisation condition involves finding the solutions in regions G and P and matching them across region E. Here we focus on region G, which is of greatest interest since it is the g-mode cavity phase integral that subsequently undergoes modification due to the magnetic field. Discussion of the remaining regions is deferred to Appendix \ref{sec:convectiveenvelope}, and relevant results from there referenced where needed.

Let us introduce the radial displacement, $X$, and the spheroidal and toroidal stream functions, $S$ and $\Theta$, respectively. Then the components of the fluid displacement vector $\boldsymbol{\xi}$ are
\begin{align}
  \xi_r &= X \:, \nonumber \\
  \xi_\theta &= \frac{1}{r \sin \theta} \frac{\del \Theta}{\del \phi} + \frac{1}{r} \frac{\del S}{\del \theta} \:, \label{eq:xi_compts} \\
  \xi_\phi &= -\frac{1}{r} \frac{\del \Theta}{\del \theta} + \frac{1}{r \sin \theta} \frac{\del S}{\del \phi} \:. \nonumber
\end{align}
We shall also write the oscillation equations (\ref{eq:osc-1})--(\ref{eq:osc-2}) in the slightly more compact form
\begin{align}
  X (\omega^2 - N^2) &= \frac{F}{\rho} \frac{\del}{\del r} \left( \frac{p'}{F} \right) \:, \label{eq:osc-3} \\
  p' \left[ 1 - \frac{\gamma p}{\rho} \frac{\ell(\ell+1)}{r^2 \omega^2} \right] &= -\frac{\gamma p}{r^2 F} \frac{\del}{\del r} \left( r^2 X F \right) \:, \label{eq:osc-4} \\
  \shortintertext{where}
  \frac{1}{F} \frac{\del F}{\del r} = \frac{1}{\gamma p} \frac{\del p}{\del r}
\end{align}
and we have dropped subscript `bg's with the understanding that unprimed quantities now refer to backgrounds.

\subsection{Region G}
This region corresponds to the g-mode cavity, occupying $r_i < r < r_e$ where $r_i$ and $r_e$ are the interior and exterior turning points (radial coordinates where $\omega^2 = N^2$). Away from the turning points, $N^2 \gg \omega^2$ and the wavelengths are very small compared to the scales of background variations, meaning that the WKBJ approximation can be used. However, this is not valid in the vicinity of the turning points, and so the solutions there will be treated in separate sub-sections.

We begin by recasting Equations (\ref{eq:osc-3})--(\ref{eq:osc-4}) in terms of the rescaled parameters $Y \equiv r^2 p^{1/\gamma} X$ and $W \equiv p'/p^{1/\gamma}$. Neglecting the first term on the LHS of (\ref{eq:osc-4}) under the assumption that $c_s^2 \to \infty$, these are
\begin{align}
  Y \left( \omega^2 - N^2 \right) &= \frac{r^2 p^{2/\gamma}}{\rho} \frac{\del W}{\del r} \:, \\
  W \frac{\ell(\ell+1)}{\rho \omega^2} &= p^{-2/\gamma} \frac{\del Y}{\del r} \:.
\end{align}
Eliminating $W$ to obtain a single second-order DE for Y, we get
\begin{align}
  \frac{\del^2 Y}{\del r^2} + \frac{\del}{\del r} \left( \ln \rho p^{-2/\gamma} \right) \frac{\del Y}{\del r} + \frac{\ell(\ell+1)}{r^2} \left( \frac{N^2}{\omega^2} - 1 \right) Y = 0 \:, \label{eq:G_DE}
\end{align}
whose WKBJ solution is
\begin{align}
  Y(r) &= C_0 \frac{p^{1/\gamma}}{\rho^{1/2}} \mathcal{U}^{-1/4} \cos \left( \int_{r_i}^r \mathcal{U}^{1/2}(r') \rmd r' + \nu \right) \label{eq:Gsoln} \\
  \shortintertext{where}
  \mathcal{U}(r) &\equiv \frac{\ell(\ell+1)}{r^2} \left( \frac{N^2}{\omega^2} - 1 \right) \:,
\end{align}
$C_0$ is an arbitrary scaling constant and $\nu$ is a constant phase.

\subsubsection{Matching to a solution near the origin}
We now match the solution (\ref{eq:Gsoln}) to one valid near $r=0$. We begin by noting that near $r=0$ we may write
\begin{align}
  \frac{\del}{\del r} \left( \ln \rho p^{-2/\gamma} \right) = \frac{1}{\rho}\frac{\del \rho}{\del r} - \frac{2}{\gamma}\frac{1}{p}\frac{\del p}{\del r} \rightarrow \frac{r}{\bar r^2} \:,
\end{align}
where ${\bar r}$ is a characteristic core radius, and
\begin{align}
  N^2 \rightarrow \frac{(N^2)^{''}}{2} r^2 \:.
\end{align}
Here the double prime denotes the second radial derivative. Thus near $r=0$, Equation (\ref{eq:G_DE}) becomes
\begin{align}
  &\frac{\del^2 Y}{\del r^2} + \frac{r}{\bar r^2} \frac{\del Y}{\del r} + \ell(\ell+1) \left( \frac{1}{q^2 {\bar r}^2} - \frac{1}{r^2} \right) Y = 0 \:, \label{eq:G_DE1} \\
  \shortintertext{where}
  &q^2 \equiv \frac{2\omega^2}{(N^2)^{''}{\bar r}^2} \:.
\end{align}
In the limit of small $\omega$, $q$ is a small parameter. Defining a new coordinate $\sigma \equiv r/(q {\bar r})$, Equation (\ref{eq:G_DE1}) becomes
\begin{align}
  \frac{\del^2 Y}{\del \sigma^2} + \sigma q^2 \frac{\del Y}{\del \sigma} + \ell(\ell+1) \left( 1 - \frac{1}{\sigma^2} \right) Y = 0 \:. \label{eq:G_DE11}
\end{align}
In the limit $q \rightarrow 0$, the solution of (\ref{eq:G_DE11}) that remains regular as $\sigma \rightarrow 0$ is to within an arbitrary multiplicative constant
\begin{align}
  Y &= \sigma^{1/2} J_{\ell+1/2}\left( \sqrt{\ell(\ell+1)} \sigma \right) \equiv \left( \frac{r}{q {\bar r}} \right)^{1/2} J_{\ell+1/2}\left( \sqrt{\ell(\ell+1)} \frac{r}{q {\bar r}} \right) \:.
\end{align}
This takes the asymptotic form
\begin{align}
  Y = \left( \frac{2}{\pi \sqrt{\ell(\ell+1)}} \right)^{1/2} \cos\left( \sqrt{\ell(\ell+1)} \frac{r}{q {\bar r}} - \frac{\pi}{2} (\ell + 1) \right) \:, \label{eq:G_DE2}
\end{align}
which, given the approximations under which (\ref{eq:G_DE2}) was derived, may also be written
\begin{align}
  Y = \left( \frac{2}{\pi \sqrt{\ell(\ell+1)}} \right)^{1/2} \cos\left( \int_{r_i}^r \mathcal{U}^{1/2}(r') \rmd r' - \frac{\pi}{2} (\ell + 1) \right) \:. \label{eq:G_DE3}
\end{align}
Comparing the phase with that of (\ref{eq:Gsoln}), we identify $\nu = -(\ell+1) \pi/2$.

\subsubsection{Solution in the vicinity of the outer turning point}\label{Gturn}
If $\ell$ is not large, then a small scale of variation justifying a WKBJ solution does not occur, because $|N^2|/\omega^2 \leq 1$ in the region $r \geq r_e$. Analytic solutions must therefore be obtained through other means of simplification.

In the vicinity of the outer turning point (where $\omega^2 = N^2$) we make the local approximation
\begin{align}
  N^2 \approx \omega^2 + (N^2)'_{r_e} (r - r_e) \:. \label{eq:N2_local}
\end{align}
Rescaling $\mathcal{Y} \equiv \rho^{1/2} p^{-1/\gamma} Y$ and defining
\begin{align}
  y \equiv (r_e - r) \left( \frac{\ell(\ell+1)}{\omega^2 r_e^2} |(N^2)'_{r_e}| \right)^{1/3} \:,
\end{align}
where we note that $(N^2)'_{r_{e}} < 0$ and consideration is restricted to $y >0$, Equation (\ref{eq:G_DE}) then becomes
\begin{align}
  \frac{\del^2 \mathcal{Y}}{\del y^2} + y \mathcal{Y} = 0 \:. \label{eq:Airy}
\end{align}

Equation (\ref{eq:Airy}) is Airy's equation which has the solution
\begin{align}
  \mathcal{Y}(y) &= C_1 y^{1/2} \left[ J_{1/3}\left( \frac{2}{3} y^{3/2} \right) + J_{-1/3}\left( \frac{2}{3} y^{3/2} \right) \right] \nonumber \\
&\quad + C_2 y^{1/2} \left[ J_{1/3}\left( \frac{2}{3} y^{3/2} \right) - J_{-1/3}\left( \frac{2}{3} y^{3/2}\right) \right] \:, \label{eq:Gsoln_TP}
\end{align}
where $J_n$  denotes the Bessel function of order $n$. We remark that choice of the linear combination of Bessel functions multiplying $C_1$ leads to a purely exponentially decaying solution in the limit where the WKBJ approximation applies in the evanescent region, $y < 0,$  while that multiplying $C_2$ yields an independent exponentially increasing solution. The asymptotic form of (\ref{eq:Gsoln_TP}) for large $y$ is given by
\begin{align}
  Y(y) = \sqrt{\frac{3}{\pi}} \frac{p^{1/\gamma}}{\rho^{1/2}} y^{-1/4} \left[\sqrt{3} C_1 \cos \left( \frac{2}{3} y^{3/2} - \frac{\pi}{4} \right) \right. \nonumber \\
    \left. - C_2 \cos\left( \frac{2}{3} y^{3/2} +\frac{\pi}{4} \right) \right] \:. \label{eq:Gsoln_TP_asym}
\end{align}
To match this with the WKBJ solution given by (\ref{eq:Gsoln}), we note that under the approximation (\ref{eq:N2_local}),
\begin{align}
  \mathcal{U}(r) \approx \frac{\ell(\ell+1)}{\omega^2 r_{e}^2} |(N^2)'_{r_{e}}| ( r_{e}-r) \\
  \shortintertext{and consequently}\int_r^{r_e} \mathcal{U}^{1/2}(r') \rmd r' \approx \frac{2}{3} {y}^{3/2} \:.
\end{align}
Accordingly near $r=r_e$, Equation (\ref{eq:Gsoln}) takes the form
\begin{align}
    Y(y) &= C_0 \frac{p^{1/\gamma}}{\rho^{1/2}} \left( \left[ \frac{\ell(\ell+1)}{\omega^2 r_e^2} |(N^2)'_{r_e}| \right]^{2/3} y \right)^{-1/4} \nonumber \\
  &\quad \times \cos \left( \frac{2}{3} y^{3/2} - \int_{r_i}^{r_e} \mathcal{U}^{1/2}(r') \rmd r' - \nu \right) \:. \label{pap1}
\end{align}
In order to match the phase of (\ref{eq:Gsoln_TP_asym}) with that of (\ref{pap1}), it must be that
\begin{align}
  \int_{r_i}^{r_e} \mathcal{U}^{1/2}(r') \rmd r' +\nu - \frac{\pi}{4} = \tan^{-1} \left( \frac{C_2}{\sqrt{3}C_1} \right) \:,
 \label{eq:nu} 
\end{align}
or equivalently
\begin{align}
  \frac{C_2}{C_1} = -\sqrt{3}\cot\left( \int_{r_i}^{r_e} \mathcal{U}^{1/2}(r') \rmd r' - \frac{\ell\pi}{2} - \frac{\pi}{4} \right) \:. \label{C8toC91}
\end{align}
Multiplicative factors for the solutions are then readily matched subsequently.

\subsection{Matching to the solution in the convective envelope}
The outer convective envelope supports p-modes that can be described using a WKBJ approximation provided that the dimensionless quantity $\omega^2 r^2/[\ell(\ell+1) c_s^2]$ is large. This is expected to be the case in the outer regions of the convective envelope below the superadiabatic zone. However, this will not be the case in the lower regions where $\omega^2$ may approach $\ell(\ell+1) c_s^2/r^2$. A breakdown of WKBJ may also occur near the outer edge of the radiative core, where $N^2/\omega^2$ approaches unity. In between there will be an evanescent zone separating the outer region where p-modes propagate from the inner region where g-modes propagate.

In order to perform the matching process, we continue the solution given by (\ref{eq:Gsoln_TP_asym}) into the outer envelope and match it to a WKBJ solution that is valid there. This leads to an additional condition on the ratio $C_2/C_1$ that takes a similar form to (\ref{C8toC91}). This can be written in the form (see Appendix \ref{sec:convectiveenvelope})
\begin{align}
  \frac{C_2}{C_1} &= \frac{A_1}{A_2}\times \left[ \sin(\nu_2 - \nu_1) \cot\left(  \int_{r_t}^{R} \Lambda^{1/2}(r) \rmd r + \nu_0 + \nu_2 \right) \right. \nonumber \\
  &\quad \left. - \cos(\nu_2 - \nu_1) \right] \:, \label{C8toC92}
\end{align}
where $\Lambda$ is defined in Equation (\ref{eq:Lambda}). Under slow background variations, this may be approximated by
\begin{align}
\Lambda \approx \frac{\omega^2 \rho}{\gamma p} - \frac{\ell(\ell+1)}{r^2} \:.
\end{align}
Here $\nu_1$ and $\nu_2$ are phases associated with the WKBJ solutions that are continuations of the two independent solutions comprising (\ref{eq:Gsoln_TP_asym}) into the convective envelope, while $A_1$ and $A_2$ are multiplicative scaling constants associated with them [see Equations (\ref{eq:Gmatch_C8})--(\ref{eq:Gmatch_C9})], and $\nu_0$ is the phase required to satisfy the surface boundary condition (see Appendix \ref{sec:regionP_surf}). These parameters in general have to be determined by numerical integration. However, analytical values can be found in the limit that a scale separation applies enabling a WKBJ approximation to be valid. This scenario is discussed further in Section \ref{sec:discuss_WKBJ}.

\subsection{Derivation of a condition determining the eigenfrequencies}
The condition determining the eigenfrequencies is obtained by equating the two expressions (\ref{C8toC91}) and (\ref{C8toC92}) for the ratio
$C_2/C_1$. The result is
\begin{align}
  &\sqrt{3}\cot\left(  \int_{r_i}^{r_e} \mathcal{U}^{1/2}(r) \rmd r - \frac{\ell\pi}{2} - \frac{\pi}{4} \right) = \frac{A_1}{A_2} \times \nonumber \\
  &\left[ \cos(\nu_1 - \nu_2) + \sin(\nu_1 - \nu_2) \cot\left( \int_{r_t}^{R} \Lambda^{1/2}(r) \rmd r + \nu_0 + \nu_2 \right) \right] \:. \label{C8toC93}
\end{align}

\section{Incorporating a core field}\label{sec:Bfield}
One effect of a magnetic field is to modify the wavenumber at a given frequency $\omega$. Although in principle the spherical symmetry of the star will also be broken, allowing for more complicated effects such as coupling between the different spherical harmonics and the excitation of torsional motions, these effects are expected to be small for fields not in the dynamically significant regime \citep{Loi2017, Loi2018}. We will incorporate the effects of a core magnetic field by considering the first-order modification to the phase integral on the LHS of Equation (\ref{C8toC93}) due to the field. That is, we seek expressions for the magnetic contribution to $\mathcal{U}^{1/2}$, which is effectively the radial wavenumber. We will neglect torsional motions and the coupling between spherical harmonics, and then determine the radial dependence of the oscillations which we will show are governed by a fourth-order system of ODEs. These may be solved numerically to yield a set of magnetically affected eigenfunctions and eigenfrequencies for given $\ell$ and $m$. Another result, that follows straightforwardly once the wavenumber modification is derived, is an expression for the frequency shift in terms of parameters relating to the field.

The benefit of this approach over perturbation theory is that the wavenumber modification can be obtained relatively easily. In perturbation theory, the correction to the eigenmodes is obtained as a superposition of basis functions, and so extracting the associated wavenumber would be far less straightforward. At higher field strengths and larger frequency shifts, where degenerate perturbation theory might become appropriate, this would be even more complicated owing to the need to solve a matrix inversion problem. However, we avoid this here.

\subsection{Equations of motion}\label{sec:eom}
We assume small-amplitude (linearised), time-harmonic motion, no rotation, and an axisymmetric field described by the poloidal flux function, $\psi$. With these assumptions, and in addition making the Cowling approximation, the equation of motion for a mode oscillating at frequency $\omega$ is
\begin{align}
  -\omega^2 \boldsymbol{\xi} &= -\frac{\nabla p'}{\rho} + \frac{\rho'}{\rho^2} \nabla p + \mathbf{F}_L \:, \label{eq:mag-eom} \\
  \shortintertext{where}
  \mathbf{F}_L &\equiv -\frac{\rho'}{\rho^2} (\nabla \times \mathbf{B}) \times \mathbf{B} + \frac{1}{\rho} \left[ (\nabla \times \mathbf{B}') \times \mathbf{B} + (\nabla \times \mathbf{B}) \times \mathbf{B}' \right] \label{eq:F_L}
\end{align}
(we are in units such that $\mu_0 = 1$). As mentioned above, we will ignore torsional motions and their associated stream function $\Theta$ (see Equation (\ref{eq:xi_compts})), since these result in higher-order contributions \citep{Loi2017}. The separate equations of motion for $S$ and $X$ can be isolated by taking the horizontal divergence and radial component of Equation (\ref{eq:mag-eom}), respectively, yielding
\begin{align}
  \omega^2 \nabla_\perp^2 S &= \frac{1}{\rho} \nabla_\perp^2 p' - \frac{r}{\sin \theta} \left( \frac{\del}{\del \theta} [\sin \theta F_{L\theta}] + \frac{\del F_{L\phi}}{\del \phi} \right) \:, \label{eq:S-eom} \\
  \omega^2 X &= \frac{1}{\rho} \frac{\del p'}{\del r} - \frac{\rho'}{\rho^2} \frac{\del p}{\del r} - F_{Lr} \:. \label{eq:X-eom}
\end{align}
Now in the limit of modes with small radial scales, the dominant terms in $\mathbf{F}_L$ are those having the highest order derivatives of $\boldsymbol{\xi}$ with respect to radius (in this case, second), since $\del^2/\del r^2 \sim k_r^2$. This allows simplifications to the full expressions for $F_{Lr}$, $F_{L\theta}$ and $F_{L\phi}$ in terms of $\psi$, $B_\phi$, $S$ and $X$ by retaining only these dominant terms, which are written out in Appendix \ref{sec:lorentz-terms}. Substituting these into (\ref{eq:S-eom})--(\ref{eq:X-eom}), we end up with
\begin{align}
  \omega^2 \nabla_\perp^2 S &= \frac{1}{\rho} \nabla_\perp^2 p' - \frac{1}{\rho r^2 \sin^2 \theta} \left( \sin \theta \frac{\del}{\del \theta} \left[ \frac{1}{\sin \theta} \frac{\del \psi}{\del \theta} \frac{\del \psi}{\del r} \right] \right. \nonumber \\
  &\qquad \left. + \frac{\del \psi}{\del \theta} \frac{\del \psi}{\del r} \frac{\del}{\del \theta} - rB_\phi \frac{\del \psi}{\del \theta} \frac{\del}{\del \phi} \right) \frac{\del^2 X}{\del r^2} \nonumber \\
  &\qquad - \frac{1}{\rho r^4 \sin^2 \theta} \left( \sin \theta \frac{\del}{\del \theta} \left[ \frac{1}{\sin \theta} \left( \frac{\del \psi}{\del \theta} \right)^2 \right] \frac{\del}{\del \theta} \right. \nonumber \\
  &\qquad \left. + \left( \frac{\del \psi}{\del \theta} \right)^2 \frac{\del^2}{\del \theta^2} + \frac{1}{\sin^2 \theta} \left( \frac{\del \psi}{\del \theta} \right)^2 \frac{\del^2}{\del \phi^2} \right) \frac{\del^2 S}{\del r^2} \:, \label{eq:S-eom-2}
\end{align}
\begin{align}
  -(\omega^2 - N^2) X &+ \frac{p^{1/\gamma}}{\rho} \frac{\del}{\del r} \left( \frac{p'}{p^{1/\gamma}} \right) = \nonumber \\
  \frac{1}{\rho} &\left[ B_\phi^2 + \frac{1}{r^2 \sin^2 \theta} \left( \frac{\del \psi}{\del r} \right)^2 \right] \frac{\del^2 X}{\del r^2} \nonumber \\
  &+ \frac{1}{\rho r^3 \sin^2 \theta} \frac{\del \psi}{\del \theta} \left[ \frac{1}{r} \frac{\del \psi}{\del r} \frac{\del}{\del \theta} - B_\phi \frac{\del}{\del \phi} \right] \frac{\del^2 S}{\del r^2} \:, \label{eq:X-eom-2}
\end{align}
where $\rho'$ has been eliminated in favour of $p'$ and $N^2$ using
\begin{align}
  \rho' = p' \frac{\rho}{\gamma p} - \rho^2 \left( \frac{\del p}{\del r} \right)^{-1} N^2 X \:.
\end{align}
The neglect of lower $r$-derivatives is justified for low-degree modes, whose smallest spatial scales occur in the radial direction.

Adopting the spherical harmonic expansion
\begin{align}
  Q = \sum_{\ell = m}^\infty Q_\ell Y_\ell^m(\theta, \phi) \:, \quad Q \in \left\{ S, X, p' \right\} \:, \label{eq:SH-expansion} \\
  \shortintertext{with the normalisation}
  \int_0^\pi Y_{\ell'}^m Y_\ell^m \sin \theta \rmd \theta =
  \begin{cases}
    1 & \text{if} \; \ell = \ell' \\
    0 & \text{if} \; \ell \neq \ell' \:,
  \end{cases}
\end{align}
these equations become
\begin{align}
  \omega^2 S_\ell &= \frac{p_\ell'}{\rho} + \frac{1}{\rho r^2 \ell(\ell+1)} \left[ \sum_{\ell'} \frac{\del^2 X_{\ell'}}{\del r^2} C_{\ell,\ell'} \right] \nonumber \\
  &\qquad + \frac{1}{\rho r^4 \ell(\ell+1)} \left[ \sum_{\ell'} \frac{\del^2 S_{\ell'}}{\del r^2} D_{\ell,\ell'} \right] \:, \label{eq:S-osc-1}
\end{align}
\begin{align}
  -(\omega^2 - N^2) X_\ell + \frac{p^{1/\gamma}}{\rho} \frac{\del}{\del r} \left[ \frac{p_\ell'}{p^{1/\gamma}} \right] &= \nonumber \\
  \frac{1}{\rho r^2} \left[ \sum_{\ell'} \frac{\del^2 X_{\ell'}}{\del r^2} E_{\ell,\ell'} \right] &+ \frac{1}{\rho r^4} \left[ \sum_{\ell'} \frac{\del^2 S_{\ell'}}{\del r^2} F_{\ell,\ell'} \right] \:, \label{eq:X-osc-1}
\end{align}
where $C_{\ell,\ell'}$, $D_{\ell,\ell'}$, $E_{\ell,\ell'}$ and $F_{\ell,\ell'}$ are coefficients describing the structure of the magnetic field. The expressions for these for a general axisymmetric field are written out in Appendix \ref{sec:mag-coeffs}. Applying the expansion (\ref{eq:SH-expansion}) to the continuity equation and adiabatic condition and combining them, we obtain
\begin{align}
  p_\ell' - \ell(\ell+1) \frac{\gamma p}{r^2} S_\ell = -\frac{\gamma p^{1-1/\gamma}}{r^2} \frac{\del (r^2 p^{1/\gamma} X_\ell)}{\del r}
\end{align}
which allows us to eliminate $p_\ell'$ in favour of $S_\ell$ in Equations (\ref{eq:S-osc-1})--(\ref{eq:X-osc-1}), giving
\begin{align}
  \left( \omega^2 - \frac{\ell(\ell+1)}{r^2} \frac{\gamma p}{\rho} \right) S_\ell + \frac{\gamma p^{1-1/\gamma}}{\rho r^2} \frac{\del}{\del r} (r^2 p^{1/\gamma} X_\ell) = \nonumber \\
  \qquad + \frac{1}{\rho r^2 \ell(\ell+1)} \left( \frac{\del^2 X_\ell}{\del r^2} C_{\ell,\ell} + \frac{1}{r^2} \frac{\del^2 S_\ell}{\del r^2} D_{\ell,\ell} \right) \:, \label{eq:S-osc-2}
\end{align}
\begin{align}
  (\omega^2 - N^2) X_\ell &+ \frac{1}{\rho r^2} \left( \frac{\del^2 X_\ell}{\del r^2} E_{\ell,\ell} + \frac{1}{r^2} \frac{\del^2 S_\ell}{\del r^2} F_{\ell,\ell} \right) = \nonumber \\
  \frac{\gamma p^{1/\gamma}}{\rho} &\frac{\del}{\del r} \left[ \ell(\ell+1) \frac{p^{1-1/\gamma}}{r^2} S_\ell - \frac{p^{1-2/\gamma}}{r^2} \frac{\del}{\del r} \left( r^2 p^{1/\gamma} X_\ell \right) \right] \:. \label{eq:X-osc-2}
\end{align}
Here we have neglected cross-coupling between different $\ell$. Notice that this is a fourth-order system despite having made the Cowling approximation (which reduces the ordinary hydrodynamic oscillation equations from fourth-order to second-order), since additional derivatives arise from the Lorentz force term. In the special case of the Prendergast solution, which has equatorial reflection symmetry, the magnetic coefficients simplify to
\begin{align}
  C_{\ell,\ell} &= 2\Psi \Psi' \int_0^\pi \left( \left[ 2 \cos^2 \theta - \sin^2 \theta \right] Y_\ell^m \right. \nonumber \\
  &\qquad \left. + \sin \theta \cos \theta \frac{\del Y_\ell^m}{\del \theta} \right) Y_\ell^{m*} \sin \theta \rmd \theta \label{eq:Prend-C} \\
  D_{\ell,\ell} &= 4\Psi^2 \int_0^\pi \left( \left[ \cos^2 \theta - 2\sin^2 \theta \right] \frac{\del Y_\ell^m}{\del \theta} \right. \nonumber \\
  &\qquad \left. + \sin \theta \cos \theta \frac{\del^2 Y_\ell^m}{\del \theta^2} - m^2 \cot \theta Y_\ell^m \right) Y_\ell^{m*} \cos \theta \rmd \theta \label{eq:Prend-D} \\
  E_{\ell,\ell} &= \Psi'^2 + \lambda^2 \Psi^2 \int_0^\pi Y_\ell^m Y_\ell^{m*} \sin^3 \theta \rmd \theta \label{eq:Prend-E} \\
  F_{\ell,\ell} &= \Psi \Psi' \int_0^\pi \sin 2\theta \frac{\del Y_\ell^m}{\del \theta} Y_\ell^{m*} \rmd \theta \:. \label{eq:Prend-F}
\end{align}

\begin{figure}
  \centering
  \includegraphics[width=\columnwidth]{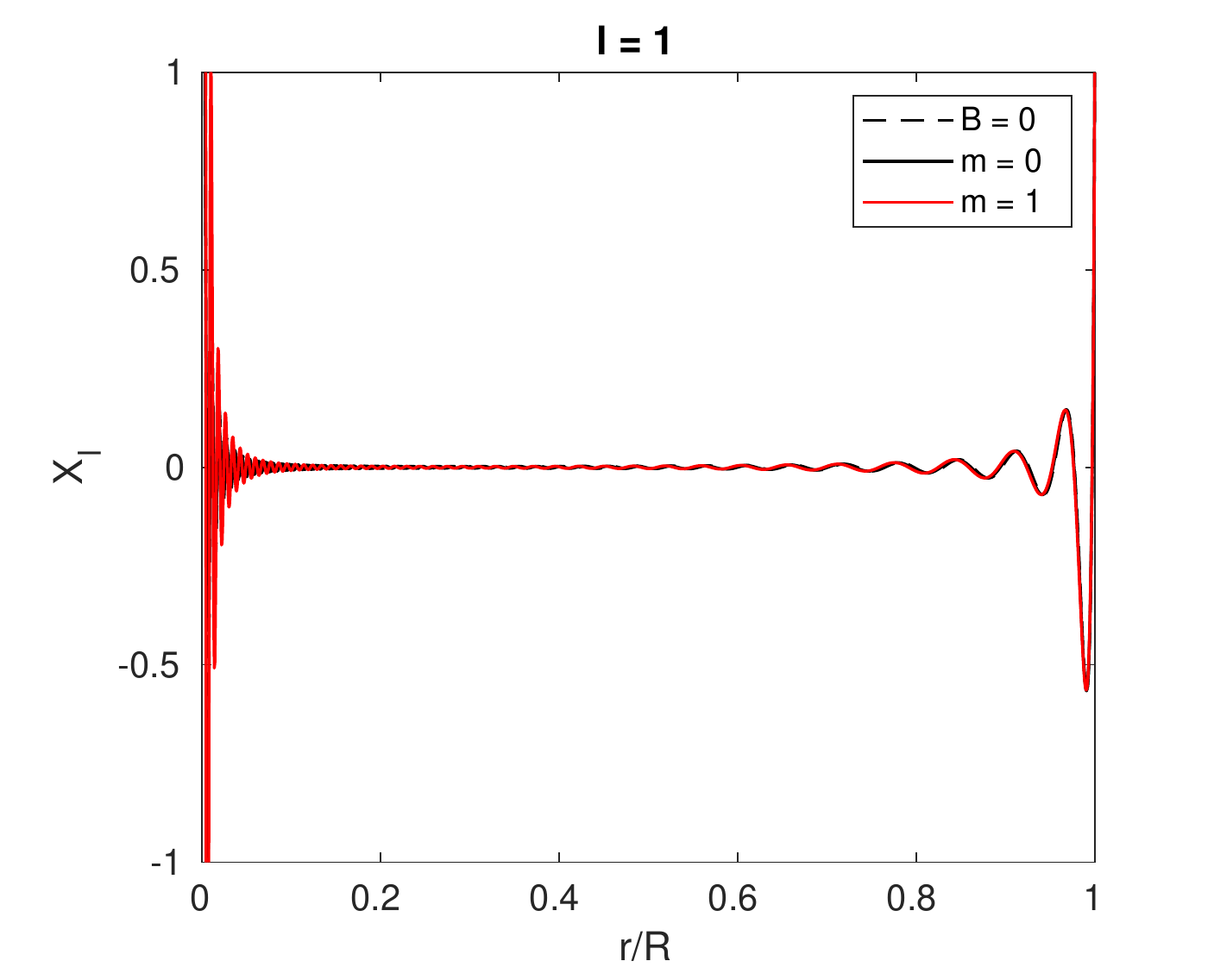}
  \includegraphics[width=\columnwidth]{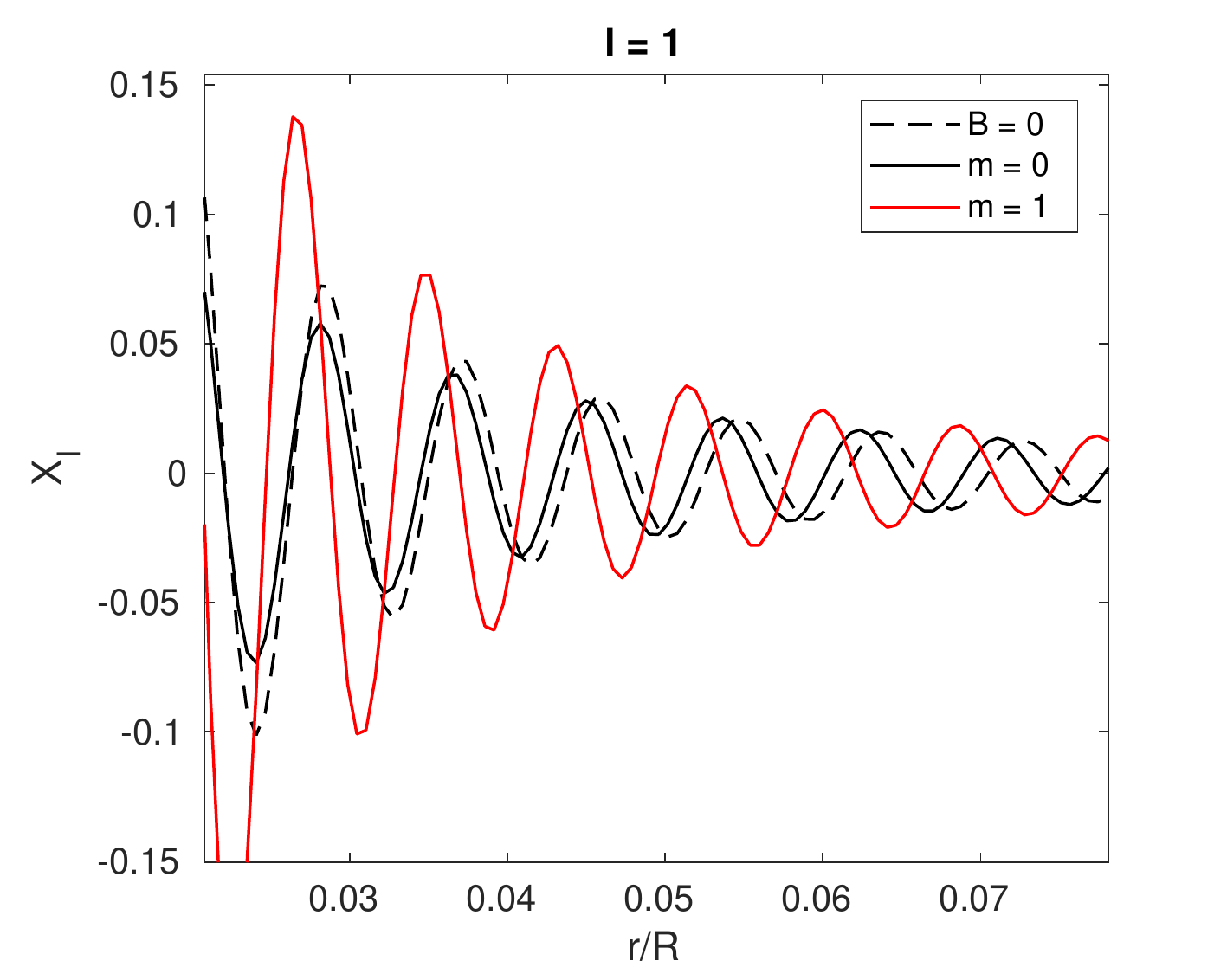}
  \caption{A selection of eigenfunctions with $n = -90$ ($\omega_0 = 0.0443$) and $m = 0, 1$, computed by numerically solving Equations (\ref{eq:S-osc-3})--(\ref{eq:X-osc-3}) for a polytrope of index 3 with a Prendergast field within $0.3 R$. The field strength is scaled such that the central Alfv\'{e}n speed is $3 \times 10^{-5} \sqrt{GM/R}$. The lower panel is a close-up showing the slight differences between the eigenfunctions of different $m$.}
  \label{fig:nonptb-mag-l1_eigs}
\end{figure}

Without a magnetic field, (\ref{eq:S-osc-2})--(\ref{eq:X-osc-2}) would reduce to a second-order system of differential equations. If the magnetic terms (which contribute the third- and fourth-order derivatives) are small compared to the hydrodynamic terms, which would be the case if the field is not dynamically significant, then the equations can be approximated by a second order system where the magnetic field contributes a small correction to the coefficients. That is, for sufficiently weak fields, Equations (\ref{eq:S-osc-2})--(\ref{eq:X-osc-2}) can be recast in the form
\begin{align}
  \frac{\del S_\ell}{\del r} &= \left( -\rho N^2 \left( \frac{\del p}{\del r} \right)^{-1} + \mathcal{M}_1 \right) S_\ell + \left( 1 - \frac{N^2}{\omega^2} + \mathcal{M}_2 \right) X_\ell \:, \label{eq:S-osc-3} \\
  \frac{\del X_\ell}{\del r} &= \left( \frac{\ell(\ell+1)}{r^2} - \frac{\rho \omega^2}{\gamma p} + \mathcal{M}_3 \right) S_\ell + \left( -\frac{2}{r} - \frac{1}{\gamma p} \frac{\del p}{\del r} + \mathcal{M}_4 \right) X_\ell \:, \label{eq:X-osc-3}
\end{align}
where the $\mathcal{M}$ coefficients, which contain the leading-order corrections associated with the magnetic field, are given in Appendix \ref{sec:M-coeffs}. Obtaining this form of the equations involves repeatedly differentiating and substituting to obtain the non-magnetic forms of the second, third and fourth derivatives of $S_\ell$ and $X_\ell$ in terms of lower-order derivatives, and then substituting these into Equations (\ref{eq:S-osc-2})--(\ref{eq:X-osc-2}) to eliminate all but the zeroth and first-order derivatives. This process retains only the leading-order magnetic terms. The resulting Equations (\ref{eq:S-osc-3})--(\ref{eq:X-osc-3}) have identical structure to the usual equations of oscillation (for which $\mathcal{M}_1 = \mathcal{M}_2 = \mathcal{M}_3 = \mathcal{M}_4 = 0$) and can be solved via identical numerical means to obtain a spectrum of eigenmodes for given $(\ell,m)$. See that this formalism allows the modes to be computed from scratch, rather than from a superposition of the unperturbed basis. A selection of these for radial order $n = -90$ and $\ell = 1$ is shown in Figure \ref{fig:nonptb-mag-l1_eigs}, along with the eigenfunction for zero field, for a polytropic model of index 3 and assuming $\gamma = 5/3$. The non-magnetic mode has a frequency of $\omega_0 = 0.0443 \sqrt{GM/R^3}$, corresponding to a low-frequency, high radial-order g-mode.

\subsection{Modification to eigenfrequency condition}\label{sec:modification_eigfreqcond}
Suppose that the eigenfrequency condition (\ref{C8toC93}) is satisfied for some resonant mode with frequency $\omega$ in the absence of a field. When the field is switched on, it modifies the wavenumber and thus perturbs the value of the phase integral. The efficiency with which this occurs depends on the strength of the field. To maintain the overall cumulative phase and therefore the resonant properties of the mode, the frequency must be altered accordingly. In this section, we will derive the expression for the wavenumber perturbation under the assumption that the magnetic field is confined to the g-mode cavity, and that the phase integral here varies much more rapidly with frequency compared to the p-mode cavity.

We begin by eliminating $S_\ell$ from (\ref{eq:S-osc-2})--(\ref{eq:X-osc-2}) to arrive at a single fourth-order ODE for $X_\ell$. For relatively weak fields, we may neglect terms quadratic or higher in powers of the magnetic coefficients $C_{\ell,\ell}$, $D_{\ell,\ell}$ etc. This yields
\begin{align}
  \alpha_1 X_\ell'''' + \alpha_2 X_\ell''' + \alpha_3 X_\ell'' + \alpha_4 X_\ell' + \alpha_5 X_\ell = 0 \:, \label{eq:X-osc-quartic}
\end{align}
where the coefficients $\alpha_1$--$\alpha_5$ are written out in Appendix \ref{sec:alpha-coeffs}. While $\alpha_1$ and $\alpha_2$ are proportional to the square of the magnetic field, and are therefore zero when the field is zero, the remaining $\alpha$ coefficients are each a sum of magnetic and non-magnetic parts. Let us write them in the form
\begin{align}
  \alpha_i = \alpha_{i0} + \alpha_{i1} \:, \quad i = 1, \cdots, 5
\end{align}
where the subscript `0' term contains only non-magnetic terms, and all magnetic terms are gathered in the subscript `1' term. In this case, $\alpha_{10} = \alpha_{20} = 0$. For weak fields, we also have $|\alpha_{i1}| \ll |\alpha_{i0}|$. Equation (\ref{eq:X-osc-quartic}) can be turned into an algebraic equation with the WKBJ ansatz $X_\ell = a(r) \exp[\rmi \int k_r \rmd r]$, where $k_r(r)$ is the (large) radial wavenumber and variations in $a(r)$ are comparatively slow. This yields
\begin{align}
  \alpha_1 k_r^4 - \rmi \alpha_2 k_r^3 - \alpha_3 k_r^2 + \rmi \alpha_4 k_r + \alpha_5 = 0 \:. \label{eq:X-osc-alg-1}
\end{align}
Let us write $k_r(r) = k_{r0}(r) + k_{r1}(r)$, where $k_{r0}$ describes the solution in the absence of a field and $k_{r1}$ represents the correction associated with the field. We shall assume $|k_{r1}| \ll k_{r0}$ (weak fields). Substituting this into (\ref{eq:X-osc-alg-1}) and performing a binomial expansion, we get
\begin{align}
  k_{r1} \approx -\frac{\alpha_{11} k_{r0}^4 - \rmi \alpha_{21} k_{r0}^3 - \alpha_{31} k_{r0}^2 + \rmi \alpha_{41} k_{r0} + \alpha_{51}}{4\alpha_1 k_{r0}^3 - 3\rmi \alpha_2 k_{r0}^2 - 2\alpha_3 k_{r0} + \rmi \alpha_4} \:, \label{eq:X-osc-alg-2}
\end{align}
where we have used the fact that $-\alpha_{30} k_{r0}^2 + \rmi \alpha_{40} k_{r0} + \alpha_{50} = 0$. The dominant term in the numerator of (\ref{eq:X-osc-alg-2}) is that with the highest power of $k_{r0}$. For the denominator, it can be shown that in the limit $c_s^2 \to \infty$
\begin{align}
  \alpha_1 &\sim \frac{B^2}{\rho} \frac{r^2}{\ell(\ell+1)} \:,\quad \alpha_2 \sim \frac{B^2}{\rho} \frac{r}{\ell(\ell+1)} \:, \nonumber \\
  \alpha_3 &\sim \frac{r^2 \omega^2}{\ell(\ell+1)} \:,\quad \alpha_4 \sim \frac{\gamma p}{\rho r} \:, \\
  \shortintertext{and so we have}
  \frac{\alpha_1}{\alpha_3} &\sim \frac{v_A^2}{\omega^2} \ll k_{r0}^{-2} \:, \\
  \frac{\alpha_2}{\alpha_4} &\sim \frac{v_A^2}{c_s^2 \ell(\ell+1)/r^2} \ll \frac{v_A^2}{\omega^2} \ll k_{r0}^{-2} \:,
\end{align}
where we make use of the fact that $k_{r0} v_A \ll \omega$ for weak fields. The dominant terms are thus the last two. For small-scale modes, the larger one of these will be $-2\alpha_3 k_{r0}$. Approximating $\alpha_3$ by its non-magnetic version, and neglecting all but the largest term in the numerator, we have approximately
\begin{align}
  k_{r1} \approx \frac{\alpha_{11} k_{r0}^3}{2\alpha_{30}} \:, \label{eq:k_r1}
\end{align}
where
\begin{align}
  \alpha_{11} &= \frac{D_{\ell,\ell}}{\rho r^6} \left( \frac{\gamma p}{\rho} \right)^2 \left[ \omega^2 - \frac{\ell(\ell+1)}{r^2} \frac{\gamma p}{\rho} \right]^{-2} \\
  \shortintertext{and}
  \alpha_{30} &= \left[ \frac{\rho}{\gamma p} - \frac{\ell(\ell+1)}{r^2 \omega^2} \right]^{-1} \:.
\end{align}
In the limit of large sound speeds, we then have
\begin{align}
  k_{r1} \approx -\frac{1}{2\omega^2} \frac{D_{\ell,\ell}}{\rho r^4} \frac{k_{r0}^3}{\ell(\ell+1)} \:.
\end{align}
Note that purely imaginary corrections to this implied by (\ref{eq:X-osc-alg-2}) are expected to vanish as $k_{r0} \to \infty$. On account of this, such corrections are not dealt with in the approximation scheme used. Under the assumptions of small $\ell$ and $\omega^2 \ll N^2$, we have
\begin{align}
  k_{r0} &\approx \frac{\sqrt{\ell(\ell+1)}}{r} \frac{N}{\omega} \\
  \shortintertext{and so for a Prendergast field}
  k_{r1} &\approx -\frac{\Psi^2 \eta \sqrt{\ell(\ell+1)} N^3}{\omega^5 \rho r^7} \:,
\end{align}
where
\begin{align}
  \eta &\equiv 2 \int_0^\pi \left[ \left( \cos^2 \theta - 2 \sin^2 \theta \right) \frac{\del Y_\ell^m}{\del \theta} + \sin \theta \cos \theta \frac{\del^2 Y_\ell^m}{\del \theta^2} \right. \nonumber \\
    &\quad \left. - m^2 \cot \theta Y_\ell^m \right] Y_\ell^{m*} \cos \theta \rmd \theta \:.
\end{align}
The eigenfrequency condition (\ref{C8toC93}) would accordingly be modified by replacing $\mathcal{U}^{1/2} \to \mathcal{U}^{1/2} + k_{r1}$, to get
\begin{align}
  &\sqrt{3}\cot\left( \int_{r_i}^{r_e} \mathcal{U}^{1/2}(r) \rmd r - \frac{\eta \sqrt{\ell(\ell+1)}}{\omega^5} \int_{r_i}^{r_e} \frac{\Psi^2 N^3}{\rho r^7} \rmd r \right. \nonumber \\
  &\quad - \left. \frac{\ell\pi}{2} - \frac{\pi}{4} \right) = \frac{A_1}{A_2} \left[ \cos(\nu_1 - \nu_2) \right. \nonumber \\
  &\qquad \left. + \sin(\nu_1 - \nu_2) \cot\left( \int_{r_t}^{R} \Lambda^{1/2}(r) \rmd r + \nu_0 + \nu_2 \right) \right] \:. \label{eq:eigfreqcond-mag}
\end{align}
Here we have assumed that the field is confined within the g-mode cavity ($r < r_e$), and we are neglecting the contribution from the region $r < r_i$ which is reasonable on account of its evanescence. Thus the leading order effect of a core magnetic field is to contribute an additional phase to the g-mode cavity with a frequency dependence going as $\omega^{-5}$.

Note that while the above specialises to a Prendergast solution, the result for an arbitrary axisymmetric field can be obtained by replacing $\eta \Psi^2$ with $D_{\ell,\ell}/2$, where the more general expression for $D_{\ell,\ell'}$ is given in Appendix \ref{sec:mag-coeffs}.

\subsection{Frequency shift}
The quantisation condition can be restated
\begin{align}
  \int_{r_i}^{r_e} \mathcal{U}^{1/2}(r) \rmd r - \frac{\eta \sqrt{\ell(\ell+1)}}{\omega^5} \int_{r_i}^{r_e} \frac{\Psi^2 N^3}{\rho r^7} \rmd r = (n + \epsilon) \pi \:,
\end{align}
where $n \in \mathbb{Z}$ is the radial order. Under the assumption that g-mode cavity resonances are much more closely spaced than those of the p-mode cavity, the phase $\epsilon$ will be a slow function of frequency, and may be regarded as nearly a constant over the frequency range between adjacent g-modes. Without a magnetic field, the second term on the LHS vanishes and the unperturbed frequency can be expressed
\begin{align}
  \omega_0 = \frac{\sqrt{\ell(\ell+1)}}{(n + \epsilon) \pi} \int_{r_i}^{r_e} \frac{N}{r} \rmd r \:.
\end{align}
Expanding $\omega = \omega_0 + \omega_1$ with $|\omega_1| \ll \omega_0$, it can be shown in the non-dynamically significant regime ($k_{0r} v_A \ll \omega_0$) that the frequency shift is given by
\begin{align}
  \omega_1 = -\frac{1}{\omega_0^4} \frac{\eta \sqrt{\ell(\ell+1)}}{(n + \epsilon) \pi} \int_{r_i}^{r_e} \frac{\Psi^2 N^3}{\rho r^7} \rmd r \:. \label{eq:freqshift}
\end{align}
While the above is for the special case of a Prendergast solution, the corresponding expression for an arbitrary axisymmetric field can be obtained by replacing $\eta \Psi^2$ above by $D_{\ell,\ell}/2$, where the general form of $D_{\ell,\ell'}$ is contained in Appendix \ref{sec:mag-coeffs}.

\begin{figure*}
  \centering
  \includegraphics[width=\textwidth]{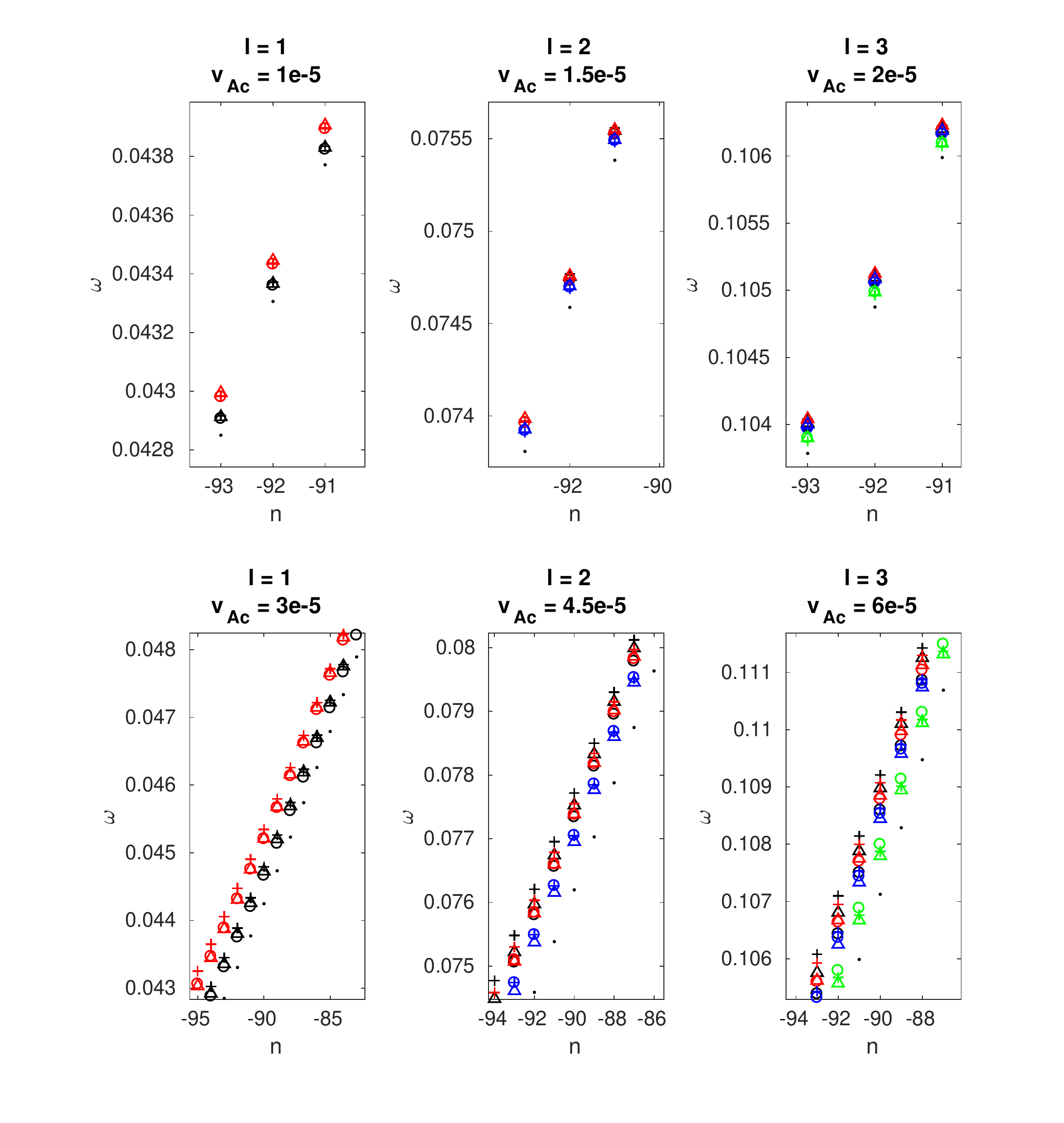}
  \caption{A comparison of mode frequencies computed via three different approaches, for a magnetised polytrope of index 3 with a Prendergast field contained within the central $0.3 R$, for various $\ell$ values and field strengths. These are indicated in the title of each panel, where $v_{Ac}$ refers to the central Alfv\'{e}n speed, expressed in units of $\sqrt{GM/R}$. Different colours correspond to different $m$ values, where $m = 0, 1, 2, 3$ are displayed in black, red, blue and green, respectively. Circle symbols correspond to numerical values computed by solving Equations (\ref{eq:S-osc-2})--(\ref{eq:X-osc-2}), while $+$ symbols correspond to analytical values given by Equation (\ref{eq:freqshift}). Solid points show the frequencies in the case of zero field. Triangle symbols correspond to the predictions of first-order non-degenerate perturbation theory, discussed in Section \ref{sec:discussion}. In the top row of panels, which are for weaker fields and smaller shifts, shown is a close-up of a small number of modes so that the frequency shifts can be seen.}
  \label{fig:nonptb-mag-l123_freqs}
\end{figure*}

A comparison between the mode frequencies predicted by (\ref{eq:freqshift}) and the numerically computed frequencies obtained by solving the system (\ref{eq:S-osc-2})--(\ref{eq:X-osc-2}) is shown in Figure \ref{fig:nonptb-mag-l123_freqs} for a selection of modes of various $\ell$ and central Alfv\'{e}n speeds. The `+' symbols were obtained as the sum of $\omega_1$ and the numerically computed frequency under zero field (black dots), while the `$\circ$' symbols were computed directly by solving (\ref{eq:S-osc-2})--(\ref{eq:X-osc-2}) for chosen $(\ell,m)$. It can be seen that the analytic formula appears to give results closely consistent with numerical values. Note that the strength of the field here is large enough to induce frequency shifts considerably larger than the spacings between g-modes of consecutive $n$.

Another point to note is that the background model is a simple polytrope that does not support mixed modes; the modes used in these calculations are g-modes. However, it still allows for a valid comparison between the methods, because an underlying assumption here is that the field is wholly confined within the g-mode cavity. We are seeking to quantify the effect of the magnetic field on the g-like portion of the mode, which is the only part assumed to be affected, and so a polytropic model supporting g-modes over its whole volume suffices for our purposes.

\section{Discussion}\label{sec:discussion}
\subsection{Comparison with perturbation theory}\label{sec:discuss-ptb}
The formalism presented here follows a considerably different approach to perturbation theory in how it treats the effects of magnetic fields on the eigenmodes and their frequencies. A pertinent test would be to check that the frequency shifts predicted by (\ref{eq:freqshift}) are consistent with those of non-degenerate perturbation theory. To briefly summarise non-degenerate perturbation theory: for small-amplitude, time-harmonic disturbances the equation of motion can be written in the general form $\mathcal{L}[\boldsymbol{\xi}_j] = \omega_j^2 \boldsymbol{\xi}_j$, where $j$ labels a particular eigenmode. Each quantity is then expanded as
\begin{align}
  \mathcal{L} &\to \mathcal{L}^{(0)} + \mathcal{L}^{(1)} + \cdots \nonumber \\
  \omega_j &\to \omega_j^{(0)} + \omega_j^{(1)} + \cdots \nonumber \\
  \boldsymbol{\xi}_j &\to \boldsymbol{\xi}_j^{(0)} + \boldsymbol{\xi}_j^{(1)} + \cdots
\end{align}
where `(0)' labels the unperturbed component, assumed to dominate, and `(1)' is the first-order correction. Using the fact that the eigenvectors of $\mathcal{L}^{(0)}$ form a complete orthonormal basis and satisfy the unperturbed equation of motion, it can be shown that the frequency shift associated with the perturbation $\mathcal{L}^{(1)}$ is
\begin{align}
  \omega_j^{(1)} = -\frac{\langle \mathcal{L}^{(1)}[\boldsymbol{\xi}_j^{(0)}], \boldsymbol{\xi}_j^{(0)} \rangle}{2\omega_j^{(0)} \langle \boldsymbol{\xi}_j^{(0)}, \boldsymbol{\xi}_j^{(0)} \rangle} \:, \label{eq:ptb-freqshift}
\end{align}
where the inner product is defined to be $\langle \boldsymbol{\xi}_i, \boldsymbol{\xi}_j \rangle \equiv \int \rho \boldsymbol{\xi}_i \cdot \boldsymbol{\xi}_j^* \rmd V$, and that the perturbation to the eigenfunction, expanded in terms of the basis of unperturbed eigenfunctions, is
\begin{align}
  \boldsymbol{\xi}_j^{(1)} &= \sum_k c_{jk} \boldsymbol{\xi}_k^{(0)} \:, \label{eq:ptb-expansion} \\
  \shortintertext{where}
  c_{jk} &= \frac{\langle \mathcal{L}^{(1)} [\boldsymbol{\xi}_j^{(0)}], \boldsymbol{\xi}_k^{(0)} \rangle}{(\omega_j^{(0)2} - \omega_k^{(0)2}) \langle \boldsymbol{\xi}_k^{(0)}, \boldsymbol{\xi}_k^{(0)} \rangle} \:. \label{eq:c_jk}
\end{align}

For ideal MHD, the operator $\mathcal{L}$ is known to be self-adjoint \citep{Freidberg2014}, and so all eigenvalues $\omega^2$ are real. Under stable background configurations, which we will assume here, it is the case that $\omega^2 > 0$ and so all frequencies $\omega$ and any frequency shifts associated with modifications to $\mathcal{L}$ will be real. On this note, one needs to be careful about the functional form of $\mathcal{L}^{(1)}$ used to calculate $\omega_j^{(1)}$. A na\"{i}ve choice might be to use the expression for $\mathbf{F}_L$ in (\ref{eq:F_L}), and set $\mathcal{L}^{(0)}$ equal to the remaining terms on the RHS of Equation (\ref{eq:mag-eom}). However, it turns out that $\mathbf{F}_L$ is not self-adjoint, which artificially introduces imaginary components into $\omega_j^{(1)}$ that are not physical. The reason this occurs is that the full effects of imposing a magnetic field include modifications to $p$ and $\rho$, which in principle need to be included in $\mathcal{L}^{(1)}$ to ensure that this remains self-adjoint. In practice, calculating the effects on the equilibrium structure is a non-trivial task and difficult to achieve analytically, so we shall not attempt that here. However, in the limit of large $c_s^2$ and short wavelengths, the term that dominates in (\ref{eq:F_L}) is $[(\nabla \times \mathbf{B}') \times \mathbf{B}]/\rho$, which conveniently turns out to be self-adjoint. Setting $\mathcal{L}^{(1)}$ to this, we are left to calculate
\begin{align}
  \omega^{(1)} &= -\frac{\int[(\nabla \times \mathbf{B}') \times \mathbf{B}] \cdot \boldsymbol{\xi}^{(0)*} \rmd V}{2\omega^{(0)} \int \rho |\boldsymbol{\xi}^{(0)}|^2 \rmd V} \nonumber \\
  &= \frac{1}{2\omega^{(0)}} \frac{\iint |\mathbf{B}'|^2 r^2 \sin \theta \rmd r \rmd \theta}{\int \rho \left[ r^2 X_\ell^{(0)2} + \ell(\ell+1) S_\ell^{(0)2} \right] \rmd r} \:, \label{eq:ptb-freqshift-mag}
\end{align}
where $\mathbf{B}'$ above is to be evaluated for the unperturbed eigenmode $\boldsymbol{\xi}^{(0)}$. The `$\triangle$' symbols in Figure \ref{fig:nonptb-mag-l123_freqs} show the predictions of perturbation theory, for comparison with the non-perturbative numerical and analytical values (`$\circ$' and `$+$' symbols). These are taken to be the sum of $\omega^{(1)}$ given by (\ref{eq:ptb-freqshift-mag}) and the numerically computed values under zero field (black dots). It appears that all three methods give similar results. We note that this quantitative agreement persists even up to moderate field strengths, where non-degenerate perturbation theory should strictly speaking no longer be used. This agreement can be regarded as somewhat fortuitous. When frequency shifts begin to exceed mode spacings, the underlying framework of the theory breaks down: perturbed modes may end up closer to a different unperturbed mode than the ``starting'' one, and so a perturbed mode can no longer be exclusively associated with a single unperturbed one. To consider this further, suppose that $|\mathcal{L}^{(1)}|/|\mathcal{L}^{(0)}|$ when applied to the functions we consider is characterised by a small parameter, $\varepsilon$. When this is large enough, such that the ratio of the unperturbed frequency separation between neighbouring modes to their actual unperturbed frequency is also of order $\varepsilon$, convergence of the series constructing the resultant eigenfunctions in Equation (\ref{eq:ptb-expansion}) becomes problematic, signifying that the choice of single unperturbed eigenfunctions as basis is inappropriate. In that situation, degenerate perturbation theory should instead be employed, which involves expanding the eigenfunctions as superpositions of neighbouring unperturbed modes to construct a new basis. This avoids the mathematical difficulties of the non-degenerate theory. It is also necessary to do this if there are avoided crossings, which could potentially occur in first-order perturbation theory in this situation, because perturbations to the mode frequencies are large enough to produce such crossings. In either case, perturbation theory does not afford a straightforward means of extracting an expression for the wavenumber modification, which was needed in deriving the magnetised eigenfrequency condition (\ref{eq:eigfreqcond-mag}). We thus have little reason to apply it here. We remark that our method, which consists of solving Equations (\ref{eq:S-osc-3}) and (\ref{eq:X-osc-3}), contains the effects of the magnetic field correct to first order and can be considered to encompass degenerate perturbation theory using an arbitrary basis (for fixed $\ell$), since construction of the final eigenfunction through numerical integration circumvents having to specify any set of basis functions. In addition, the eigenfrequency condition (\ref{eq:eigfreqcond-mag}) is obtained from these equations using a WKBJ analysis, which enables calculation of the modified wavenumber as the gradient of the phase function in the usual way. We therefore do not expect this approach to be significantly more or less accurate than standard first-order methods, since this aspect of the calculation closely parallels pre-existing approaches.

As a comment, we note that our use of the term ``non-perturbative'' stems from an attempt to highlight the distinction between our method and standard perturbation methods. On a philosophical level, however, our approach may still be regarded as perturbative in a sense. The difference is that the perturbations are taken with respect to the governing equations of motion (since we retain only leading terms in the Lorentz force and neglect cross-$\ell$ coupling, etc.) and not from the point of view of the eigenfunctions and eigenfrequencies, which is the case for standard perturbative methods. In doing so, we are freed from some of the limitations posed by the latter, and have a larger parameter space of applicability compared to non-degenerate theory. Take note that this primarily relates to the ability to represent the resultant eigenfunction, from which the perturbed wavenumber might in principle be obtained, rather than the ability to compute the frequency shift, which appears not to suffer in the moderate regime. The neglect of higher-order Lorentz terms still limits the range of field strengths for which our method is valid; in particular, it is not expected to hold up to the dynamically significant regime (where $\omega_A \sim \omega$). Having emphasised this, we hope that our use of ``non-perturbative'' shall not be misleading in that regard.

\subsection{Relevance for seismic inversion studies}
The main result of this paper, which is the criterion for magnetised mode formation given by Equation (\ref{eq:eigfreqcond-mag}), simultaneously ties together a large number of parameters describing a star, including properties of the g-mode cavity (through the integral over $\mathcal{U}^{1/2}$), the parameters of a possible core magnetic field (through the term involving the weighted integral over $\Psi^2$), the structure of the evanescent region (through the coefficients $A_1, A_2$ and phases $\nu_1, \nu_2$), and the properties of the p-mode cavity (through the integral over $\Lambda^{1/2}$). Qualitatively, if the g-mode phase integral varies much more rapidly than the p-mode phase integral, then in the case of no magnetic field ($\Psi = 0$), Equation (\ref{eq:eigfreqcond-mag}) predicts a spectrum of modes that are roughly equally spaced in period. The spacing of these modes is modulated over periodic intervals corresponding to the dynamical frequency of the star (the p-mode spacing, i.e.~the large separation), and the shape of the modulation pattern is controlled by parameters relating to the evanescent zone. Some further insight can be gained if we rearrange Equation (\ref{eq:eigfreqcond-mag}) into the more lucid form
\begin{align}
  (\cot K_g - \Sigma_g) \tan K_p = \frac{A_1}{\sqrt{3} A_2} \sin(\nu_1 - \nu_2) \:, \label{eq:eigfreqcond2}
\end{align}
where
\begin{align}
  K_p &\equiv \int_{r_t}^R \Lambda^{1/2}(r) \rmd r + \nu_0 + \nu_2 \:, \label{eq:Kp} \\
  K_g &\equiv \int_{r_i}^{r_e} \mathcal{U}^{1/2}(r) \rmd r - \frac{\ell\pi}{2} - \frac{\pi}{4} - \frac{\eta\sqrt{\ell(\ell+1)}}{\omega^5} \int_{r_i}^{r_e} \frac{\Psi^2 N^3}{\rho r^7} \rmd r \:, \label{eq:Kg}
\end{align}
\begin{align}
  \Sigma_g &\equiv \frac{A_1}{\sqrt{3} A_2} \cos(\nu_1 - \nu_2) \:. \label{eq:Sigma_g}
\end{align}
The offset term $\Sigma_g$ is expected to be a slow function of frequency and can be regarded as roughly a constant. In the limit of a thick evanescent zone, $A_1/A_2 \to 0$ and we would obtain the separate conditions $\tan K_p = 0$ and $\cot K_g = 0$, which correspond to the pure p-mode and g-mode eigenspectra.

It can be seen that the functional form of Equation (\ref{eq:eigfreqcond2}) differs slightly from the familiar form seen in previous works \citep{Shibahashi1979, Tassoul1980, Takata2016a}, which is
\begin{align}
  \cot \Theta_g \tan \Theta_p = q_c \label{eq:eigfreqcond-lit}
\end{align}
where $\Theta_g$ and $\Theta_p$ are the g-mode and p-mode cavity phase integrals (akin to $K_g$ and $K_p$ above), and $q_c$ is a coupling parameter relating to the evanescent zone. In our case, we obtain an extra offset $\Sigma_g$ in the g-mode factor. Although it is not algebraically possible to absorb an offset into the argument of a cot function and thereby get rid of $\Sigma_g$ above, we note that Equations (\ref{eq:eigfreqcond2}) and (\ref{eq:eigfreqcond-lit}) would still effectively produce spectra with identical properties. This is because the condition $\cot K_g = \Sigma_g$ is satisfied whenever $K_g$ changes by $\pi$ units, regardless of the value of $\Sigma_g$, and hence the g-mode spacings would be unaffected. It would act to contribute a global frequency offset to the spectrum. Since it becomes negligible when $A_1/A_2$ tends to zero, it would appear that the physical origin of $\Sigma_g$ lies in the non-WKBJ nature of strongly coupled cavities. This is discussed more in Section \ref{sec:discuss_WKBJ}.

An advancement over previous works is the inclusion of a possible magnetic field within the g-mode cavity, where the field is assumed not to be strong enough to affect gravity wave propagation significantly (and thus not induce additional effects such as mode damping or chaos), but strong enough to lie beyond the scope of validity of non-degenerate perturbation theory. Our formalism contrasts perturbation theory in that an expression for the modification to the wavenumber (and thereby the phase integral) can be straightforwardly obtained, since the magnetically-affected eigenfunctions are themselves solutions to a fourth-order system of ODEs, rather than being considered a superposition of an unperturbed basis. This allows the magnetic terms to be incorporated explicitly in the criterion for the mode frequencies. Following directly on from the wavenumber modification is an expression for the frequency shift, which is given in Equation (\ref{eq:freqshift}). The quantity $\eta$ and the weighted integral over $\Psi^2$ capture information about the latitudinal and radial structure of the magnetic field, whose parametrisation we have simplified by assuming a Prendergast (twisted-torus) solution, believed to be a realistic configuration for a confined magnetic equilibrium. Such a solution is described by a single scalar function $\Psi(r)$, and so despite being for a three-dimensional structure, the inversion problem would still be a purely radial one. We also demonstrate how the properties of the evanescent region might be captured in the eigenfrequency condition for the case where analytic treatments may not work properly in this region, but where a numerical integration instead might be required. This is likely to be the case for low-degree modes and thin evanescent zones.

In the case where the field is a more general axisymmetric configuration, the results (\ref{eq:eigfreqcond-mag}) and (\ref{eq:freqshift}) would remain valid but with a modification to replace $\eta \Psi^2$ with $D_{\ell,\ell}/2$ as described. The difference would be that the radial and latitudinal dependencies do not separate, and the inversion problem would be more complicated.

\subsection{The WKBJ limit for high-degree modes}\label{sec:discuss_WKBJ}
Matching of the solutions from region G, which are given by (\ref{eq:Gmatch_C8}) and (\ref{eq:Gmatch_C9}), across E and into P, is achieved through adjustment of the amplitude coefficients $A_1, A_2$ and phases $\nu_1, \nu_2$ as described in Appendix \ref{sec:convectiveenvelope}. For a simple power-law model of the lower convective envelope, the solutions are a linear combination of $J_{\pm\tau}(2\alpha \zeta^{1/2})$, where $\zeta$ is a scaled radial coordinate and $\tau$, $\alpha$ are parameters describing the background. For definitions of $\tau$, $\alpha$ and $\zeta$, as well as details of the model, see Appendix \ref{sec:regionP_rt}. The requirements to match (\ref{eq:Gmatch_C8}) to the asymptotic form of $J_{-\tau}(2\alpha \zeta^{1/2})$ and (\ref{eq:Gmatch_C9}) to that of $J_\tau(2\alpha \zeta^{1/2})$ [see Equation (\ref{eq:Lsoln_asym})] are that $\nu_1 = \tau\pi - \pi/4 - \pi\tau'/2$ and $\nu_2 = -\pi/2 - \pi\tau'/2$, where
\begin{align}
  \tau' \equiv \frac{1}{\beta \left(\sqrt{4\ell(\ell+1) + (k-1)^2} + \sqrt{4\ell(\ell+1) + k(k-2)}\right)} \:.
\end{align}
The quantities $\beta$, $k$ and $\tau$ characterise the background structure and are also defined in Appendix \ref{sec:regionP_rt}. To obtain these results, we made use of the fact that
\begin{align}
  2\alpha \zeta^{1/2} - \frac{\pi\tau}{2}- \int^r_{r_t} \Lambda^{1/2}(r') dr' \rightarrow - \frac{\pi\tau'}{2} \label{eq:phase}
\end{align}
as $\zeta \to \infty$. See that $\tau' \to 0$ as $\ell \to \infty$, which means that in this limit $\nu_1 \to \tau\pi - \pi/4$ and $\nu_2 \to -\pi/4$. The eigenfrequency condition then becomes
\begin{align}
  &\sqrt{3} \cot\left( \int_{r_i}^{r_e} \mathcal{U}^{1/2}(r) \rmd r - \frac{\ell\pi}{2} - \frac{\pi}{4} \right) = \frac{A_1}{A_2} \times \nonumber \\
  &\quad \left[ \cos(\pi\tau) + \sin(\pi\tau) \cot\left( \int_{r_t}^R \Lambda^{1/2}(r) \rmd r + \nu_0 - \frac{\pi}{4} \right) \right] \:. \label{C8toC94}
\end{align}

The quantity $|A_1/A_2|$, by construction, measures the ratio of magnitudes of the exponentially decreasing and exponentially increasing solutions on the P side of the evanescent zone. For large $\ell$, this will be exponentially small and dominated by the factor $\exp(-2\int |k_e| \rmd r)$, where
\begin{align}
  |k_e| = \sqrt{\left( 1 - \frac{N^2}{\omega^2} \right) \left( \frac{\ell(\ell+1)}{r^2} - \frac{\rho\omega^2}{\gamma p} \right)}
\end{align}
and the integral is taken over the evanescent region at the boundaries of which $k_e$ vanishes. In this situation the first cosine term on the RHS of (\ref{C8toC94}) may be dropped in comparison to the LHS, which may be balanced on the RHS by a large value of the cotangent term able to compensate for the exponentially small multiplicative factor. Thus in the full WKBJ limit we get
\begin{align}
  &\sqrt{3}\cot\left( \int_{r_i}^{r_e} \mathcal{U}^{1/2}(r) \rmd r - \frac{\ell\pi}{2} - \frac{\pi}{4} \right) = \frac{A_1}{A_2} \times \nonumber \\
  &\quad \sin(\pi\tau) \cot \left( \int_{r_t}^R \Lambda^{1/2}(r) \rmd r + \nu_0 - \frac{\pi}{4} \right) \:, \label{C8toC95}
\end{align}
which is of the form of (\ref{eq:eigfreqcond-lit}), the mixed-mode quantisation condition familiar from the literature. In this limit, the extra offset $\Sigma_g$ present in (\ref{eq:eigfreqcond2}) does not appear. These simplifications occur for large $\ell$. For the case of small $\ell$, a large exponential amplitude growth or decay may not be expected in the evanescent zone and the phases $\nu_1$ and $\nu_2$ may take on different values. These are sensitive to the structure of the evanescent zone and lower part of the p-mode cavity. Measurements of $\Sigma_g$ could in principle provide constraints on these phases, thereby yielding information about the properties of these regions.

As a final remark on the full WKBJ limit, we note that the phase $\nu_1$ was obtained from connecting an exponentially decreasing solution into a propagation zone. However, contamination by an exponentially increasing solution that becomes negligible interior to the turning point makes such a determination imprecise. We here indicate that the form of (\ref{C8toC95}) is unaffected by this consideration. Noting that the $J_\tau$ term becomes small interior to the evanescent region, suppose we matched to (\ref{eq:Lsoln_asym1}) rather than use (\ref{eq:Lsoln_asym}) to determine $\nu_1$. Then we would conclude that $\nu_1 = \pi/4 - \pi\tau'/2 \to \pi/4$. Furthermore, the factor $\sin(\pi\tau)$ in (\ref{eq:Lsoln_asym1}) implies that the amplitude factor $A_1 \rightarrow A_1\sin(\pi\tau)$. Adopting these prescriptions in (\ref{C8toC93}), we recover (\ref{C8toC95}).

\subsection{Limitations}
A number of assumptions and simplifications have been made to obtain the results in this paper, which have already been mentioned in preceding sections. To summarise the main ones, these include (i) neglecting rotation, (ii) neglecting modification to the equilibrium structure by the magnetic field, since we assume spherically symmetric backgrounds, (iii) neglect of the cross-coupling between modes of different $\ell$, and (iv) ignoring the excitation of torsional motions and associated feedback. With the exception of (i), the other effects are expected to be small for magnetic fields that do not occupy the dynamically significant regime, which we do not focus on here. From a practical point of view, the applicability of our results to the seismic inversion of actual red giant power spectra is likely to be most heavily limited by our neglect of rotational effects, since these by themselves are known to produce observable frequency shifts/splittings.

It is worthy to note that the same formalism developed here for the case of magnetism, where modifications to the wavenumber and corresponding frequency shifts are treated in a non-perturbative fashion, can be extended to the case of rotation. If the rotational and magnetic axes are aligned, then inclusion of the Coriolis terms alongside the Lorentz terms in the starting equations, followed by the same spherical harmonic expansion and neglect of cross-$\ell$ coupling etc.~would add rotation terms to the phase integrals in Equation (\ref{eq:eigfreqcond-mag}). An expression for the rotational frequency shift could then be analogously derived, and the overall frequency shift for given $\ell$ and $m$ would be a sum of the magnetic and rotational contributions. The case of misaligned rotational and magnetic axes would be less straightforward to treat. Further work on this front needs to be done.

\section{Summary}\label{sec:summary}
We have developed a non-perturbative formalism (N.B.~see elaboration at the end of Section \ref{sec:discuss-ptb}) for treating the effects of magnetic fields on mixed modes in red giant stars, where the magnetic field is assumed to be confined to the radiative core. Our results for the eigenfrequency condition, given in Equation (\ref{eq:eigfreqcond-mag}), and the frequency shift, given in Equation (\ref{eq:freqshift}), are expected to hold for fields of up to moderate strengths, i.e.~those that do not occupy the dynamically significant regime (associated with disruption to wave propagation). We have verified that predicted frequency shifts are consistent with non-degenerate perturbation theory. The results derived here may be useful in seismic inversion studies; in particular if the Lorentz (magnetic) splitting can be measured, then this would provide constraints on the field strength and distribution. We also explored the possibility of non-WKBJ effects relevant to mixed modes of low degree. Our results demonstrate that the form of the quantisation condition remains very similar to that in the case where WKBJ is applicable throughout the star. However, in our more general expression, an additional offset $\Sigma_g$ appears, given in Equation (\ref{eq:Sigma_g}). This is unrelated to magnetic effects and has its origins in the non-WKBJ nature of strongly coupled cavities. Measurements of this may provide information about the structure of the evanescent zone and base of the convective envelope. The effects of rotation have not yet been incorporated and would be the subject of future work.

\section*{Acknowledgments}
STL is supported by funding from the Cambridge Australia Trust, and Churchill College, Cambridge, through a Junior Research Fellowship.

%%%%%%%%%%%%%%%%%%%%%%%%%%%%%%%%%%%%%%%%%%%%%%%%%%

%%%%%%%%%%%%%%%%%%%% REFERENCES %%%%%%%%%%%%%%%%%%

% The best way to enter references is to use BibTeX:

\bibliographystyle{mnras}
%\bibliography{C:/Users/STCLoi/Documents/articles_and_papers/my_stuff/refs}

% Alternatively you could enter them by hand, like this:
% This method is tedious and prone to error if you have lots of references

%%%%%%%%%%%%%%%%%%%%%%%%%%%%%%%%%%%%%%%%%%%%%%%%%%

%%%%%%%%%%%%%%%%% APPENDICES %%%%%%%%%%%%%%%%%%%%%

\appendix

\section{Treatment of the upper layers}\label{sec:convectiveenvelope}
In this appendix we discuss the convective envelope and evanescent zone (regions P and E), for the purposes of connecting the solutions in these regions to those in the g-mode cavity (treated in Section \ref{sec:mixedmodes}), and thereby obtaining the mixed-mode quantisation condition.

\subsection{Region P}
Here we still shall assume $N^2 = 0$, making the region isentropic. Equations (\ref{eq:osc-3})--(\ref{eq:osc-4}) then combine to give
\begin{align}
  \mathcal{W} \left[ \omega^2 - \frac{\gamma p}{\rho} \frac{\ell(\ell+1)}{r^2} \right] = -\frac{\gamma p}{r^2 \rho^2} \frac{\del}{\del r} \left( r^2 \rho \frac{\del \mathcal{W}}{\del r} \right) \label{eq:Ueq}\:,
\end{align}
where $\mathcal{W} \equiv p'/\rho$. Making a change of variable $V \equiv r \rho^{1/2} \mathcal{W}$, we obtain the DE
\begin{align}
  \frac{\del^2 V}{\del r^2} &= -\Lambda(r) V(r) \:, \label{eq:U_DE} \\
  \shortintertext{where}
  \Lambda(r) &= \frac{\omega^2 \rho}{\gamma p} - \frac{\ell(\ell+1)}{r^2} - \frac{1}{r \rho} \frac{\del \rho}{\del r} - \frac{1}{2\rho} \frac{\del^2 \rho}{\del r^2} + \frac{1}{4 \rho^2} \left( \frac{\del \rho}{\del r} \right)^2 \:. \label{eq:Lambda}
\end{align}
The first term in the expression for $\Lambda$ is expected to dominate, since this corresponds to the squared acoustic wavenumber, which greatly exceeds $1/r$ for typical $\omega$ of stochastically-driven oscillations in the outer parts of the convection zone.

Retaining only the first term in $\Lambda$ when considering amplitude factors in the WKBJ approximation, the WKBJ solution to (\ref{eq:U_DE}) is
\begin{align}
  V(r) = C_3 \frac{p^{1/4}}{\rho^{1/4}} \cos \left( \int_{r_t}^r \Lambda^{1/2}(r') \rmd r' + \eta \right) \:, \label{eq:Usoln}
\end{align}
corresponding to
\begin{align}
Y(r) = C_3 \frac{ r p^{1/\gamma - 1/4}}{\omega\sqrt{\gamma} \rho^{1/4}} \cos \left( \int_{r_t}^r \Lambda^{1/2}(r') \rmd r' + \eta+\pi/2 \right) \:, \label{eq:UWsoln}
\end{align}
where $C_3$ and $\eta$ are arbitrary constants. Here $r_t$ is the lower turning point where $\Lambda$ vanishes. We shall assume for simplicity that this exceeds $r_0$, where $N^2$ drops to zero from a positive value further in. Otherwise, the lower limit of the integral should be replaced by $r_0$, as (\ref{eq:Ueq}) does not apply where $N^2 \neq 0$. The constants $C_3$ and $\eta$ are in the first instance determined by the requirement that the solution (\ref{eq:Usoln}) is a continuation of the one that is specified just interior to $r_e$. Asserting the continuity of the solutions multiplying $C_1$ and $C_2$ in Equation (\ref{eq:Gsoln_TP}), we write
\begin{align}
  & y^{1/2} \left(J_{1/3}\left( \frac{2}{3} y^{3/2} \right) + J_{-1/3}\left( \frac{2}{3} y^{3/2} \right) \right) \nonumber \\
  &\qquad \rightarrow \frac{A_1 r p^{1/\gamma-1/4}}{\rho^{1/4}} \cos \left( \int_{r_t}^r \Lambda^{1/2}(r') \rmd r' + \nu_1 \right) \:, \label{eq:Gmatch_C8} \\
  & y^{1/2} \left(J_{1/3}\left( \frac{2}{3} y^{3/2} \right) - J_{-1/3}\left( \frac{2}{3} y^{3/2}\right)\right) \nonumber \\
  &\qquad \rightarrow \frac{A_2 r p^{1/\gamma-1/4}}{\rho^{1/4}} \cos \left( \int_{r_t}^r \Lambda^{1/2}(r') \rmd r' + \nu_2 \right) \:. \label{eq:Gmatch_C9}
\end{align}
The amplitude factors $A_1$ and $A_2$ and the phases $\nu_1$ and $\nu_2$ in general have to be determined from numerical integrations if the WKBJ approach cannot be used throughout region E, as will be the case for small $\ell$.

\subsubsection{Incorporating a surface boundary condition}\label{sec:regionP_surf}
The solution representing a mode of oscillation must also satisfy a boundary condition at the surface. Following the usual procedure when constructing WKBJ solutions, we assume that this condition can be met with a solution of the form (\ref{eq:UWsoln}) by specifying the phase $\eta = -\nu_0 - \int_{r_t}^R \Lambda^{1/2}(r) \rmd r - \pi/2$. The value of $\nu_0$ is determined by the surface boundary condition. Given the above continuations, it is readily found that the condition that the linear combination occurring in Equation (\ref{eq:Gsoln_TP}) leads to the correct phase condition specified above in this section is given by Equation (\ref{C8toC92}).

\subsubsection{Solution near the inner turning point}\label{sec:regionP_rt}
We consider the region of the convective envelope below the superadiabatic region, where we may take $N^2$ to be zero. The base of this region lies at $r = r_0$, and we assume that the p-mode cavity turning point lies at $r_t > r_0$. Defining $\rho_0 \equiv \rho(r_0)$ and $p_0 \equiv p(r_0)$, we shall assume the following functional forms for $\rho$ and $p$ in the region near $r_t$:
\begin{align}
  \rho = \rho_0 \left( \frac{r}{r_0} \right)^{-k} \:, \quad p = p_0 \left( \frac{r}{r_0} \right)^{-\gamma k} \:, \label{eq:L_powerlaws}
\end{align}
with $k$ assumed to be a constant that we are free to specify.

 Let us rescale the radial coordinate to $\chi \equiv r/r_0$. Setting $N^2 = 0$ and invoking the adiabatic condition $p^{1/\gamma}/\rho$ = const., Equations (\ref{eq:osc-3})--(\ref{eq:osc-4}) can be combined into a single second-order DE for $\mathcal{W}$:
\begin{align}
  \mathcal{W} \frac{\rho_0}{\gamma p_0} r_0^2 \omega^2 \chi^{2+k(\gamma-1)} = -\chi^2 \frac{\del^2 \mathcal{W}}{\del \chi^2} - (2-k) \chi \frac{\del \mathcal{W}}{\del \chi} + \ell(\ell+1) \mathcal{W} \:.
\end{align}
Defining the auxilliary quantity $\beta \equiv 2+k(\gamma-1)$, and rescaling the radial coordinate to $\zeta \equiv \chi^\beta$, the DE becomes
\begin{align}
  \zeta^2 \frac{\del^2 \mathcal{W}}{\del \zeta^2} &+ \frac{\beta+1-k}{\beta} \zeta \frac{\del \mathcal{W}}{\del \zeta} + \left[ -\frac{\ell(\ell+1)}{\beta^2} + \alpha^2 \zeta \right] \mathcal{W} = 0 \label{eq:L_DE} \:, \\
  \shortintertext{where}
  \alpha^2 &\equiv \frac{\rho_0}{\gamma p_0} \frac{r_0^2 \omega^2}{\beta^2} \:.
\end{align}
The general solution is
\begin{align}
  \mathcal{W}(\zeta) &= \zeta^{\frac{k-1}{2\beta}} \sqrt{\pi} \left[ C_4 J_\tau(2\alpha \zeta^{1/2}) + C_5 J_{-\tau}(2\alpha \zeta^{1/2}) \right] \:, \label{eq:Lsoln} \\
  \shortintertext{where}
  \tau &= \frac{1}{\beta} \left[ 4\ell(\ell+1) + (k-1)^2 \right]^{1/2}
\end{align}
and $C_4, C_5$ are arbitrary constants.

The phases $\nu_1$ and $\nu_2$ can be determined by extending the solutions from region G out to $r = r_t$ and then matching slopes and values with a solution of the form given by (\ref{eq:Lsoln}). Rather than follow this procedure through in detail for this simplified model, we remark that as indicated in Section \ref{Gturn} in the limit of large $\ell$ (when a WKBJ approach should be valid throughout), the solution (\ref{eq:Gmatch_C9}) is expected to increase exponentially into the evanescent zone while for (\ref{eq:Gmatch_C8}) the opposite is the case. Accordingly, we expect (\ref{eq:Gmatch_C8}) to match to the solution containing $J_{-\tau}(2\alpha \zeta^{1/2})$ and (\ref{eq:Gmatch_C9}) to match to the solution containing $J_\tau(2\alpha \zeta^{1/2})$.

For large $\zeta$ the solution approaches the asymptotic form
\begin{align}
  \mathcal{W}(\zeta) \approx \alpha^{-1/2} \zeta^{\frac{k-1}{2\beta} - \frac{1}{4}} \left[ C_4 \cos\left( 2\alpha \zeta^{1/2} - \frac{\tau\pi}{2} - \frac{\pi}{4} \right) \right. \nonumber \\
    \left. + C_5 \cos\left( 2\alpha \zeta^{1/2} + \frac{\tau\pi}{2} - \frac{\pi}{4} \right) \right] \:. \label{eq:Lsoln_asym}
\end{align}
From the above results we may also derive the asymptotic relation
\begin{align}
  &J_{-\tau}(2\alpha \zeta^{1/2}) - J_\tau(2\alpha \zeta^{1/2}) \cos(\pi\tau) \nonumber \\
 &\qquad \approx \left( \pi \alpha \sqrt{\zeta} \right)^{-1/2} \sin(\pi\tau) \cos\left( 2\alpha \zeta^{1/2} - \frac{\tau\pi}{2} + \frac{\pi}{4} \right) \:. \label{eq:Lsoln_asym1}
\end{align}

\section{Lorentz terms}\label{sec:lorentz-terms}
Here we write out the expressions of quantities relevant to Section \ref{sec:eom}, in which the magnetised equations of motion are derived. These include the components of the field perturbation $\mathbf{B}'$ and Lorentz force $\mathbf{F}_L$, which are relevant for Equations (\ref{eq:F_L})--(\ref{eq:X-eom-2}).

Using the linearised induction equation $\mathbf{B}' = \nabla \times (\boldsymbol{\xi} \times \mathbf{B})$, the components of the field perturbation can be written in terms of $\psi$, $B_\phi$ and the components of $\boldsymbol{\xi}$ as follows:
\begin{align}
  B_r' &= -\frac{1}{r \sin \theta} \left[ \frac{1}{r} \frac{\del}{\del \theta} \left( \xi_r \frac{\del \psi}{\del r} \right) + \frac{1}{r^2} \frac{\del}{\del \theta} \left( \xi_\theta \frac{\del \psi}{\del \theta} \right) \right. \nonumber \\
    &\qquad \left. + \frac{1}{r^2 \sin \theta} \frac{\del \psi}{\del \theta} \frac{\del \xi_\phi}{\del \phi} - B_\phi \frac{\del \xi_r}{\del \phi} \right] \:, \\
  B_\theta' &= \frac{1}{r \sin \theta} \left[ B_\phi \frac{\del \xi_\theta}{\del \phi} + \frac{1}{r \sin \theta} \frac{\del \xi_\phi}{\del \phi} \frac{\del \psi}{\del r} + \frac{\del}{\del r} \left( \xi_r \frac{\del \psi}{\del r} + \frac{\xi_\theta}{r} \frac{\del \psi}{\del \theta} \right) \right] \:, \\
  B_\phi' &= \frac{1}{r} \left[ \frac{1}{\sin \theta} \frac{\del \psi}{\del \theta} \frac{\del}{\del r} \left( \frac{\xi_\phi}{r} \right) - \frac{\del}{\del r} (r \xi_r B_\phi) - \frac{\del}{\del \theta} (\xi_\theta B_\phi) \right. \nonumber \\
    &\qquad \left. - \frac{1}{r} \frac{\del \psi}{\del r} \frac{\del}{\del \theta} \left( \frac{\xi_\phi}{\sin \theta} \right) \right] \:.
\end{align}
The full expressions for the components of $\mathbf{F}_L$ defined in Equation (\ref{eq:F_L}) in terms of $\psi$, $B_\phi$ and $\mathbf{B}' = (B_r', B_\theta', B_\phi')$ are lengthy, but retaining only the terms with second radial derivatives of $\boldsymbol{\xi}$, which are expected to dominate in the limit of small radial scales:
\begin{align}
  F_{Lr} &= \frac{1}{\rho} \left[ \left( B_\phi^2 - \frac{B_\theta}{r \sin \theta} \frac{\del \psi}{\del r} \right) \frac{\del^2 \xi_r}{\del r^2} - \frac{B_\theta}{r^2 \sin \theta} \frac{\del \psi}{\del \theta} \frac{\del^2 \xi_\theta}{\del r^2} \right. \nonumber \\
    &\quad \left. - \frac{B_\phi}{r^2 \sin \theta} \frac{\del \psi}{\del \theta} \frac{\del^2 \xi_\phi}{\del r^2} \right] \:, \\
  F_{L\theta} &= \frac{1}{\rho} \left[ \frac{B_r}{r \sin \theta} \frac{\del \psi}{\del r} \frac{\del^2 \xi_r}{\del r^2} + \frac{B_r}{r^2 \sin \theta} \frac{\del \psi}{\del \theta} \frac{\del^2 \xi_\theta}{\del r^2} \right] \:, \\
  F_{L\phi} &= \frac{1}{\rho} \left[ \frac{B_r}{r^2 \sin \theta} \frac{\del \psi}{\del \theta} \frac{\del^2 \xi_\phi}{\del r^2} - B_r B_\phi \frac{\del^2 \xi_r}{\del r^2} \right] \:.
\end{align}
These expressions are used to obtain (\ref{eq:S-eom-2})--(\ref{eq:X-eom-2}) following substitution into Equations (\ref{eq:S-eom})--(\ref{eq:X-eom}).

\section{Magnetic coefficients}\label{sec:mag-coeffs}
The following coefficients appear in Equations (\ref{eq:S-osc-1}) and (\ref{eq:X-osc-1}), which are the magnetised oscillation equations. They quantify the strength and structure of the field, and the degree of coupling between different spherical harmonics induced by the field. 
\begin{align}
  C_{\ell,\ell'} &= \int_0^\pi \left( \sin \theta \frac{\del}{\del \theta} \left[ \frac{1}{\sin \theta} \frac{\del \psi}{\del \theta} \frac{\del \psi}{\del r} \right] Y_{\ell'}^m + \frac{\del \psi}{\del \theta} \frac{\del \psi}{\del r} \frac{\del Y_{\ell'}^m}{\del \theta} \right. \nonumber \\
  &\qquad \left. - \rmi mrB_\phi \frac{\del \psi}{\del \theta} Y_{\ell'}^m \right) \frac{Y_\ell^{m*}}{\sin \theta} \rmd \theta \:, \\
  D_{\ell,\ell'} &= \int_0^\pi \left\{ \frac{\del}{\del \theta} \left[ \frac{1}{\sin \theta} \left( \frac{\del \psi}{\del \theta} \right)^2 \right] \frac{\del Y_{\ell'}^m}{\del \theta} + \left( \frac{\del \psi}{\del \theta} \right)^2 \frac{\del^2 Y_{\ell'}^m}{\del \theta^2} \frac{1}{\sin \theta} \right. \nonumber \\
    &\qquad \left. - m^2 \left( \frac{\del \psi}{\del \theta} \right)^2 \frac{Y_{\ell'}^m}{\sin^3 \theta} \right\} Y_\ell^{m*} \rmd \theta \:, \\
  E_{\ell,\ell'} &= \int_0^\pi \left[ r^2 B_\phi^2 + \frac{1}{\sin^2 \theta} \left( \frac{\del \psi}{\del r} \right)^2 \right] Y_{\ell'}^m Y_\ell^{m*} \sin \theta \rmd \theta \:, \\
  F_{\ell,\ell'} &= \int_0^\pi \frac{\del \psi}{\del \theta} \left( \frac{\del \psi}{\del r} \frac{\del Y_{\ell'}^m}{\del \theta} - \rmi mrB_\phi Y_{\ell'}^m \right) \frac{Y_\ell^{m*}}{\sin \theta} \rmd \theta \:.
\end{align}
Subsequent working assumes $\ell=\ell'$ and specialises to the Prendergast configuration (described in Section \ref{sec:Prendergast}), but the above are for the more general case, where $\ell$ and $\ell'$ may not be equal and the field is an arbitrary axisymmetric one described by poloidal flux function $\psi(r, \theta)$ and toroidal component $B_\phi(r, \theta)$. For a Prendergast field, $\psi(r, \theta) = \Psi(r) \sin^2 \theta$ and $B_\phi(r, \theta) = -\lambda \Psi(r) \sin \theta / r$, which simplifies these coefficients to the ones listed in Equations (\ref{eq:Prend-C})--(\ref{eq:Prend-F}).

\section{Coefficients of 2nd-order system}\label{sec:M-coeffs}
The following parameters quantify the leading-order corrections to the coefficients of the standard (non-magnetic) equations of stellar oscillation in the Cowling approximation, which form a second-order system. The modified equations containing these parameters are (\ref{eq:S-osc-3})--(\ref{eq:X-osc-3}), which are subsequently solved numerically to obtain the magnetised eigenfunctions and eigenfrequencies. At this stage, cross-$\ell$ coupling has been neglected, but the details of the field configuration are unspecified. These are determined through the coefficients $C_{\ell,\ell}$, $D_{\ell,\ell}$, $E_{\ell,\ell}$ and $F_{\ell,\ell}$, which are defined in Appendix \ref{sec:mag-coeffs} for the general axisymmetric case.
\begin{align}
  \mathcal{M}_1 &= \frac{1}{\rho r^2 \omega^2} \frac{1}{\ell(\ell+1)} \left[ C_{\ell,\ell} w_3 + \frac{1}{r^2} D_{\ell,\ell} w_1 \right. \nonumber \\
    &\quad + \left( \frac{\del D_{\ell,\ell}}{\del r} - \frac{4D_{\ell,\ell}}{r} - \frac{1}{\gamma p} \frac{\del p}{\del r} D_{\ell,\ell} + \ell(\ell+1) F_{\ell,\ell} \right) \frac{w_5}{r^2} \nonumber \\
    &\quad \left. + \left( \frac{\del C_{\ell,\ell}}{\del r} - \frac{2C_{\ell,\ell}}{r} - \frac{1}{\gamma p} \frac{\del p}{\del r} C_{\ell,\ell} + \ell(\ell+1) E_{\ell,\ell} \right) w_7 \right] \\
  \mathcal{M}_2 &= \frac{1}{\rho r^2 \omega^2} \frac{1}{\ell(\ell+1)} \left[ C_{\ell,\ell} w_4 + \frac{1}{r^2} D_{\ell,\ell} w_2 \right. \nonumber \\
    &\quad \left( \frac{\del D_{\ell,\ell}}{\del r} - \frac{4D_{\ell,\ell}}{r} - \frac{1}{\gamma p} \frac{\del p}{\del r} D_{\ell,\ell} + \ell(\ell+1) F_{\ell,\ell} \right) \frac{w_6}{r^2} \nonumber \\
    &\quad \left. \left( \frac{\del C_{\ell,\ell}}{\del r} - \frac{2C_{\ell,\ell}}{r} - \frac{1}{\gamma p} \frac{\del p}{\del r} C_{\ell,\ell} + \ell(\ell+1) E_{\ell,\ell} \right) w_8 \right]
\end{align}
\begin{align}
  \mathcal{M}_3 &= \frac{1}{\gamma p r^2} \frac{1}{\ell(\ell+1)} \left( C_{\ell,\ell} w_7 + \frac{1}{r^2} D_{\ell,\ell} w_5 \right) \\
  \mathcal{M}_4 &= \frac{1}{\gamma p r^2} \frac{1}{\ell(\ell+1)} \left( C_{\ell,\ell} w_8 + \frac{1}{r^2} D_{\ell,\ell} w_6 \right) \:,
\end{align}
where
\begin{align}
  w_1 &\equiv -\left( 1 - \frac{N^2}{\omega^2} \right) \left( \frac{2\ell(\ell+1)}{r^2} - \frac{\rho \omega^2}{\gamma p} \right) \frac{2}{r} - \frac{2}{\omega^2} \frac{\del N^2}{\del r} \frac{\ell(\ell+1)}{r^2} \nonumber \\
  &\quad - \left( 1 - \frac{N^2}{\omega^2} \right) \frac{1}{\gamma p} \frac{\del p}{\del r} \left( \frac{\ell(\ell+1)}{r^2} - \frac{\rho \omega^2}{p} \right) \nonumber \\
  &\quad -\left( \frac{\del p}{\del r} \right)^{-1} \left[ \rho N^2 \left( 1 - \frac{N^2}{\omega^2} \right) \frac{2\ell(\ell+1)}{r^2} - \rho^2 N^2 \frac{\omega^2}{\gamma p} \right. \nonumber \\
    &\qquad \left. + \frac{\del^2 \rho}{\del r^2} N^2 + \rho \frac{\del^2 N^2}{\del r^2} \right] + \left( \frac{\del p}{\del r} \right)^{-2} \left( 2\rho N^4 \frac{\del \rho}{\del r} \right. \nonumber \\
  &\qquad \left. + \rho^2 N^2 \frac{\del N^2}{\del r} + 2 \frac{\del \rho}{\del r} \frac{\del^2 p}{\del r^2} N^2 + \rho \frac{\del^3 p}{\del r^3} N^2 + 2\rho \frac{\del^2 p}{\del r^2} \frac{\del N^2}{\del r} \right) \nonumber \\
  &\quad - \rho N^2 \left( \frac{\del p}{\del r} \right)^{-3} \left[ 3\rho N^2 \frac{\del^2 p}{\del r^2} + 2 \left( \frac{\del^2 p}{\del r^2} \right)^2 \right]
\end{align}
\begin{align}
  w_2 &\equiv \left( 1 - \frac{N^2}{\omega^2} \right) \left( \frac{\del p}{\del r} \right)^2 \frac{1}{\gamma p^2} \left( 1 + \frac{1}{\gamma} \right) + \frac{2}{\gamma p} \frac{\del p}{\del r} \left[ \left( 1 - \frac{N^2}{\omega^2} \right) \frac{2}{r} \right. \nonumber \\
    &\qquad \left. + \frac{1}{\omega^2} \frac{\del N^2}{\del r} \right] + \left( 1 - \frac{N^2}{\omega^2} \right) \left( \frac{6}{r^2} - \frac{1}{\gamma p} \frac{\del^2 p}{\del r^2} \right) \nonumber \\
  &\quad + \left( 1 - \frac{N^2}{\omega^2} \right)^2 \left( \frac{\ell(\ell+1)}{r^2} - \frac{\rho \omega^2}{\gamma p} \right) + \frac{1}{\omega^2} \left( \frac{4}{r} \frac{\del N^2}{\del r} - \frac{\del^2 N^2}{\del r^2} \right) \nonumber \\
  &\quad + \left( \frac{\del p}{\del r} \right)^{-1} \left[ \left( 1 - \frac{N^2}{\omega^2} \right) \left( \frac{2}{r} \rho N^2 - \frac{\del \rho}{\del r} N^2 - 2\rho \frac{\del N^2}{\del r} \right) \right. \nonumber \\
    &\qquad \left. + \frac{\rho N^2}{\omega^2} \frac{\del N^2}{\del r} \right] + \left( \frac{\del p}{\del r} \right)^{-2} \left( 1 - \frac{N^2}{\omega^2} \right) 2\rho N^2 \frac{\del^2 p}{\del r^2}
\end{align}
\begin{align}
  w_3 &\equiv \left( \frac{\del p}{\del r} \right)^2 \left( 2 + \frac{1}{\gamma} \right) \frac{1}{\gamma p^2} \left( \frac{\ell(\ell+1)}{r^2} - \frac{\rho \omega^2}{p} \right) \nonumber \\
  &\quad + \frac{2}{r} \frac{1}{\gamma p} \frac{\del p}{\del r} \left[ \frac{4\ell(\ell+1)}{r^2} - \frac{\rho \omega^2}{p} \left( 2 + \frac{1}{\gamma} \right) \right] \nonumber \\
  &\quad + \left( 1 - \frac{N^2}{\omega^2} \right) \left( \frac{\ell(\ell+1)}{r^2} - \frac{\rho \omega^2}{\gamma p} \right)^2 \nonumber \\
  &\quad + 2 \left( \frac{\ell(\ell+1)}{r^2} - \frac{\rho \omega^2}{\gamma p} \right) \left( \frac{4}{r^2} - \frac{1}{\gamma p} \frac{\del^2 p}{\del r^2} \right) + \frac{10\ell(\ell+1)}{r^4} \nonumber \\
  &\quad + \frac{\rho \omega^2}{\gamma p} \left( \frac{\rho N^2}{\gamma p} + \frac{1}{p} \frac{\del^2 p}{\del r^2} - \frac{1}{\rho} \frac{\del^2 \rho}{\del r^2} \right) \nonumber \\
  &\quad + \rho \left( \frac{\del p}{\del r} \right)^{-1} \left[ 2N^2 \left( \frac{3\ell(\ell+1)}{r^3} + \frac{\omega^2}{\gamma p} \frac{\del \rho}{\del r} \right) \right. \nonumber \\
    &\qquad \left. - \left( \frac{\ell(\ell+1)}{r^2} - \frac{\rho \omega^2}{\gamma p} \right) \frac{\del N^2}{\del r} \right] \nonumber \\
  &\quad + \left( \frac{\del p}{\del r} \right)^{-2} \left( \frac{\ell(\ell+1)}{r^2} - \frac{\rho \omega^2}{\gamma p} \right) \rho N^2 \frac{\del^2 p}{\del r^2}
\end{align}
\begin{align}
  w_4 &\equiv -\left( \frac{\del p}{\del r} \right)^3 \frac{1}{\gamma p^3} \left( 2 + \frac{1}{\gamma} \right) \left( 1 + \frac{1}{\gamma} \right) - \frac{6}{r} \frac{1}{\gamma p^2} \left( \frac{\del p}{\del r} \right)^2 \left( 1 + \frac{1}{\gamma} \right) \nonumber \\
  &\quad - \frac{1}{\gamma p} \frac{\del p}{\del r} \left[ \left( 1 - \frac{N^2}{\omega^2} \right) \left( \frac{2\ell(\ell+1)}{r^2} - \frac{\rho \omega^2}{p} \left( 2 + \frac{1}{\gamma} \right) \right) \right. \nonumber \\
    &\qquad \left. + \frac{18}{r^2} - \frac{3}{p} \frac{\del^2 p}{\del r^2} \left( 1 + \frac{1}{\gamma} \right) \right] - \left( 1 - \frac{N^2}{\omega^2} \right) \left( \frac{8\ell(\ell+1)}{r^3} \right. \nonumber \\
  &\qquad \left. + \frac{\omega^2}{\gamma p} \left( \frac{\del \rho}{\del r} - \frac{4\rho}{r} \right) \right) + \frac{6}{r} \left( \frac{1}{\gamma p} \frac{\del^2 p}{\del r^2} - \frac{4}{r^2} \right) \nonumber \\
  &\quad - \left( \frac{\ell(\ell+1)}{r^2} - \frac{\rho \omega^2}{\gamma p} \right) \frac{1}{\omega^2} \frac{\del N^2}{\del r} - \frac{1}{\gamma p} \frac{\del^3 p}{\del r^3} \nonumber \\
  &\quad - \rho N^2 \left( 1 - \frac{N^2}{\omega^2} \right) \frac{\ell(\ell+1)}{r^2} \left( \frac{\del p}{\del r} \right)^{-1}
\end{align}
\begin{align}
  w_5 &\equiv \left( 1 - \frac{N^2}{\omega^2} \right) \frac{\ell(\ell+1)}{r^2} - \frac{\rho \omega^2}{\gamma p} - \left( \frac{\del p}{\del r} \right)^{-1} \rho \frac{\del N^2}{\del r} \nonumber \\
  &\quad + \left( \frac{\del p}{\del r} \right)^{-2} \rho N^2 \frac{\del^2 p}{\del r^2}
\end{align}
\begin{align}  
  w_6 &\equiv -\left( 1 - \frac{N^2}{\omega^2} \right) \left[ \frac{2}{r} + \frac{1}{\gamma p} \frac{\del p}{\del r} + \rho N^2 \left( \frac{\del p}{\del r} \right)^{-1} \right] - \frac{1}{\omega^2} \frac{\del N^2}{\del r}
\end{align}
\begin{align}
  w_7 &\equiv -\frac{1}{\gamma p} \frac{\del p}{\del r} \left( \frac{\ell(\ell+1)}{r^2} - \frac{\rho \omega^2}{p} \right) - \frac{2}{r} \left( \frac{\ell(\ell+1)}{r^2} - \frac{\rho \omega^2}{\gamma p} \right) \nonumber \\
  &\quad - \frac{\ell(\ell+1)}{r^2} \rho N^2 \left( \frac{\del p}{\del r} \right)^{-1}
\end{align}
\begin{align}
  w_8 &\equiv \left( \frac{\ell(\ell+1)}{r^2} - \frac{\rho \omega^2}{\gamma p} \right) \left( 1 - \frac{N^2}{\omega^2} \right) + \frac{6}{r^2} - \frac{1}{\gamma p} \frac{\del^2 p}{\del r^2} + \frac{4}{r} \frac{1}{\gamma p} \frac{\del p}{\del r} \nonumber \\
  &\quad + \frac{1}{\gamma p^2} \left( \frac{\del p}{\del r} \right)^2 \left( 1 + \frac{1}{\gamma} \right) \:.
\end{align}

\section{Coefficients of 4th-order system}\label{sec:alpha-coeffs}
The coefficients defined in this section appear in Equation (\ref{eq:X-osc-quartic}), which is a single fourth-order ODE into which the magnetised oscillation equations have been cast. They are subsequently used in deriving the expression for the leading-order modification to the wavenumber and thereby the magnetic modification to the eigenfrequency condition (see Section \ref{sec:modification_eigfreqcond}). 
\begin{align}
  \alpha_1 &\equiv \frac{\mu_1 \mu_4 \nu_2}{\mu_2^2} \\
  \alpha_2 &\equiv 2\frac{\mu_1 \mu_4}{\mu_2} \left( \frac{\nu_2}{\mu_2} \right)' + \frac{\mu_1 \mu_4 \nu_3}{\mu_2^2} + \mu_4 \left( \frac{\mu_1 \nu_2}{\mu_2^2} \right)' - \frac{\mu_4 \nu_1}{\mu_2} \nonumber \\
  &\quad - \frac{\mu_3 \nu_2}{\mu_2} + \frac{\mu_1 \mu_5 \nu_2}{\mu_2^2} \\
  \alpha_3 &\equiv 2\mu_4 \left( \frac{\mu_1}{\mu_2} \right)' \left( \frac{\nu_2}{\mu_2} \right)' + 3\frac{\mu_1 \mu_4}{\mu_2} \left( \frac{\nu_2}{\mu_2} \right)'' + \mu_4 \left( \frac{\mu_1 \nu_3}{\mu_2^2} \right)' - \frac{\mu_4 \nu_2}{\mu_2} \nonumber \\
  &\quad + 2\frac{\mu_1 \mu_4}{\mu_2} \left( \frac{\nu_3}{\mu_2} \right)' - \mu_4 \left( \frac{\nu_1}{\mu_2} \right)' - 2\mu_3 \left( \frac{\nu_2}{\mu_2} \right)' - \frac{\mu_3 \nu_3}{\mu_2} \nonumber \\
  &\quad - \frac{\mu_5 \nu_1}{\mu_2} + 2\frac{\mu_1 \mu_5}{\mu_2} \left( \frac{\nu_2}{\mu_2} \right)' + \frac{\mu_1 \mu_5 \nu_3}{\mu_2^2} + \nu_4
\end{align}
\begin{align}
  \alpha_4 &\equiv \nu_5 - \mu_3 \left( \frac{\nu_2}{\mu_2} \right)'' - 2\mu_3 \left( \frac{\nu_3}{\mu_2} \right)' - \mu_4 \left( \frac{\nu_2}{\mu_2} \right)' + \mu_4 \left( \frac{\mu_1}{\mu_2} \right)' \left( \frac{\nu_2}{\mu_2} \right)'' \nonumber \\
  &\quad + \frac{\mu_1 \mu_4}{\mu_2} \left( \frac{\nu_2}{\mu_2} \right)''' + 2\mu_4 \left( \frac{\mu_1}{\mu_2} \right)' \left( \frac{\nu_3}{\mu_2} \right)' + 3\frac{\mu_1 \mu_4}{\mu_2} \left( \frac{\nu_3}{\mu_2} \right)'' \nonumber \\
  &\quad - \frac{\mu_4 \nu_3}{\mu_2} - \frac{\mu_5 \nu_2}{\mu_2} + \frac{\mu_1 \mu_5}{\mu_2} \left( \frac{\nu_2}{\mu_2} \right)'' + 2\frac{\mu_1 \mu_5}{\mu_2} \left( \frac{\nu_3}{\mu_2} \right)' \\
  \alpha_5 &\equiv \nu_6 - \mu_3 \left( \frac{\nu_3}{\mu_2} \right)'' - \mu_4 \left( \frac{\nu_3}{\mu_2} \right)' + \mu_4 \left( \frac{\mu_1}{\mu_2} \right)' \left( \frac{\nu_3}{\mu_2} \right)'' \nonumber \\
  &\quad + \frac{\mu_1 \mu_4}{\mu_2} \left( \frac{\nu_3}{\mu_2} \right)''' - \frac{\mu_5 \nu_3}{\mu_2} + \frac{\mu_1 \mu_5}{\mu_2} \left( \frac{\nu_3}{\mu_2} \right)''
\end{align}
\begin{align}
  \mu_1 &= -\frac{D_{\ell,\ell}}{\rho r^4 \ell(\ell+1)} \\
  \mu_2 &= \omega^2 - \frac{\ell(\ell+1)}{r^2} \frac{\gamma p}{\rho} \\
  \mu_3 &= -\frac{F_{\ell,\ell}}{\rho r^4} \\
  \mu_4 &= \frac{\ell(\ell+1)}{r^2} \frac{\gamma p}{\rho} \\
  \mu_5 &= \frac{\ell(\ell+1)}{\rho r^2} \left[ (\gamma-1) \frac{\del p}{\del r} - \frac{2\gamma p}{r} \right]
\end{align}
\begin{align}
  \nu_1 &= \frac{C_{\ell,\ell}}{\rho r^2 \ell(\ell+1)} \\
  \nu_2 &= -\frac{\gamma p}{\rho} \\
  \nu_3 &= -\frac{1}{\rho} \left[ \frac{\del p}{\del r} + \frac{2\gamma p}{r} \right] \\
  \nu_4 &= \frac{1}{\rho} \left( \frac{E_{\ell,\ell}}{r^2} + \gamma p \right) \\
  \nu_5 &= \frac{\gamma}{\rho} \left( \frac{\del p}{\del r} + \frac{2p}{r} \right) \\
  \nu_6 &= \frac{1}{\rho} \left[ \frac{2(\gamma-1)}{r} \frac{\del p}{\del r} - \frac{2\gamma p}{r^2} - \frac{1}{\gamma p} \left( \frac{\del p}{\del r} \right)^2 + \frac{\del^2 p}{\del r^2} \right] + \omega^2 - N^2 \:.
\end{align}

%%%%%%%%%%%%%%%%%%%%%%%%%%%%%%%%%%%%%%%%%%%%%%%%%%

% Don't change these lines
\bsp	% typesetting comment
\label{lastpage}
\end{document}